\documentclass[superscriptaddress, amsmath,amssymb, aps, twocolumn, nofootinbib ]{revtex4-1}

\usepackage{graphicx}
\usepackage{dcolumn}
\usepackage{bm}
\usepackage{wasysym}
\usepackage{hyperref}
\usepackage[mathlines]{lineno}
\usepackage{xcolor}
\usepackage{dsfont}

\usepackage{gensymb}
\usepackage{enumitem}
\usepackage{multirow}
\usepackage{booktabs}
\usepackage{array}
\newcolumntype{P}[1]{>{\centering\arraybackslash}p{#1}}
\usepackage{amssymb}
\usepackage{adjustbox}
\usepackage{orcidlink}
\usepackage{float}

\usepackage[normalem]{ulem}

\newcommand*{\defeq}{\mathrel{\vcenter{\baselineskip0.5ex \lineskiplimit0pt
                     \hbox{\scriptsize.}\hbox{\scriptsize.}}}
                     =}

\newcommand{\iu}{\mathrm{i}\mkern1mu}
\newcommand{\du}{\mathrm{d}}

\begin{document}

\title{Long-term Stable Nonlinear Evolutions of Ultracompact Black-Hole
Mimickers}

\author{Gareth Arturo Marks \orcidlink{0009-0003-3160-9337}}
\email{gam54@cam.ac.uk}
\affiliation{DAMTP, Centre for Mathematical Sciences,
University of Cambridge, Wilberforce Road, Cambridge CB3 0WA, United Kingdom}

\author{Seppe J. Staelens \orcidlink{0000-0002-1262-1600}}
\email{ss3033@cam.ac.uk}
\affiliation{DAMTP, Centre for Mathematical Sciences,
University of Cambridge, Wilberforce Road, Cambridge CB3 0WA, UK}
\affiliation{Leuven Gravity Institute, KU Leuven,
Celestijnenlaan 200D box 2415, 3001 Leuven, Belgium}

\author{Tamara Evstafyeva \orcidlink{0000-0002-2818-701X}}
\email{tevstafyeva@perimeterinstitute.ca}
\affiliation{Perimeter Institute for Theoretical Physics,
Waterloo, Ontario N2L 2Y5, Canada}

\author{Ulrich Sperhake.
\orcidlink{0000-0002-3134-7088}}
\email{U.Sperhake@damtp.cam.ac.uk}
\affiliation{DAMTP, Centre for Mathematical Sciences,
University of Cambridge, Wilberforce Road, Cambridge CB3 0WA, UK}
\affiliation{Department of Physics and Astronomy, Johns Hopkins University,
3400 North Charles Street, Baltimore, Maryland 21218, USA}
\affiliation{{TAPIR 350-17, Caltech, 1200 E. California Boulevard,
Pasadena, California 91125, USA}}

\date{\today}

\begin{abstract}
We study the stability of ultracompact boson stars admitting light
rings combining a perturbative analysis with 3+1 numerical-relativity
simulations with and without symmetry assumptions. We observe
excellent agreement between all perturbative and numerical results
which uniformly support the hypothesis that this family of black-hole
mimickers is separated into stable and unstable branches by
extremal-mass configurations. This separation includes, in particular,
thin-shell boson stars with light rings located on the stable branch
which we conclude to represent long-term stable black-hole mimickers.
\end{abstract}

\maketitle

{\textit{\textbf{Introduction}}---}Black-hole solutions in closed analytic
form have played a crucial role in many aspects of the theoretical
understanding of general relativity (GR) almost right from the
moment Einstein published his theory. Starting with observations
of X-ray sources (e.g.~Cygnus X-1 \cite{Bowyer:1965}) and quasars
(e.g.~3C 273 \cite{Hazard:1963,Schmidt:1963}) and their eventual
identification with accretion onto compact objects, black holes
(BHs) acquired an equally prominent role as seemingly common citizens
of our Universe.  The rapidly increasing number of gravitational-wave
(GW) detections by LIGO and Virgo \cite{Abbott:2016blz,KAGRA:2021vkt},
all compatible with theoretical predictions for BHs and neutron
stars in GR \cite{LIGOScientific:2021sio}, and direct observations
of BH environments by the Event Horizon Telescope
\cite{EventHorizonTelescope:2019dse,EventHorizonTelescope:2022wkp} have
further cemented the prominence of BHs in observational astronomy.

In both mathematical modeling as well as popular imagination, the
most characteristic feature of a BH is its event horizon, the 2+1 dimensional outer boundary of the set of spacetime
points from which no null geodesics can reach future null infinity. In terms of their observational phenomenology, however,
the decisive feature of BHs is their (unstable) light rings (LRs),
circular null geodesics on which photons can orbit; for a Schwarzschild
BH of mass $M$, the event horizon is located at areal radius
$r=2GM/c^2$ and the LR at $r=3GM/c^2$. The appearance of accreting BHs is strongly influenced by the presence and properties of the LR, while quasinormal modes have been shown to have a close connection with null unstable geodesics
\cite{Synge:1966okc,Cardoso:2008bp,Jusufi:2019ltj,Koga:2022dsu,Volkel:2022khh,Pedrotti:2025idg}.
The observational evidence therefore points more directly at the
existence of LRs, not necessarily BHs. Additionally, classical black
holes come with theoretical issues, like the singularity problem \cite{Hawking:1973uf} or the
information paradox \cite{Mathur:2009ip}, motivating the study of
BH mimickers. Even if one accepts Occam's
razor and parsimoniously interprets BHs as the most likely sources,
there remains the intriguing possibility that a population of
ultracompact objects (UCOs) is hiding inside a more conventional
BH society and may provide us with clues about the dark-matter candidate
fields they are composed of. We also note that the GW emission from compact boson-star (BS) binaries
exhibits significant degeneracy with that from BH
binaries \cite{Evstafyeva:2024qvp}.

Clearly, dynamical longevity is a key requirement of any viable
alternative to the BH paradigm.  The stability of UCOs or \textit{BH
mimickers} -- both implied henceforth to denote objects with a LR
but no horizon -- has, consequently, gained much attention in recent
years. Under a broad set of assumptions, BH mimickers have been
shown to possess two LRs, one stable and one unstable located at
the minimum and maximum, respectively, of the null geodesic's
potential \cite{Cunha:2017qtt,Cunha:2020azh}; Kerr BHs, in contrast
have unstable LRs only.  The existence of a minimum in the effective
potential leads to \textit{stable trapping}, i.e.~null geodesics
originating close to the minimum may remain indefinitely bound to
it. At linear level, waves propagating on a background spacetime
with stable trapping exhibit decay slower than polynomial or
exponential as occurs for unstable trapping \cite{Keir:2014oka}.
Guided by results on the instability of anti-de Sitter spacetime
and its similar decay rates
\cite{Bizon:2011gg,Holzegel:2011uu,Holzegel:2013kna}, it has been
heuristically argued that UCOs may also be nonlinearly unstable,
thus ruling them out as plausible BH mimickers \cite{Cardoso:2014sna}.
In the absence of rigorous analytic results, however, numerical
evolutions are the only available tool for testing this hypothesis.

To date, literature results investigating the ultimate fate and
stability of UCOs are inconclusive. By simulating two families of
ultracompact BSs, Ref.~\cite{Cunha:2022gde} finds either migration
to nonultracompact configurations or collapse to a BH, typically
on timescales $\lesssim \mathcal{O}(10^3)\,\mu^{-1}$ in units of
the scalar mass parameter, and concludes that the LRs trigger
instability. Reference \cite{Cunha:2025oeu} explains the instability
as the result of the backreaction of perturbations at the stable
LR appearing to deepen the geodesic potential.  In contrast,
simulations of the merger and ringdown of ultracompact spinning BSs
yield no evidence of this instability \cite{Siemonsen:2024snb},
although the timescales the simulations span in that work are
considerably shorter.  Further simulations in spherical symmetry
of scalarized Reisner-Nordstr{\"o}m BHs with LRs also exhibited
long-term stability \cite{Guo:2024cts}.  In a recent study using a
toy-model spacetime with two LRs, Ref.~\cite{Benomio:2024lev}
analyzes the differences between linear and noninear dynamics in
more detail: within the region of stable trapping, the slow linear
decay leads to a growth of higher-order derivatives and energy in
the nonlinear evolution, thereby resulting in turbulent behaviour.
This phenomenon is proven in the perturbative regime
\cite{Redondo-Yuste:2025hlv}, where the authors argue how it may
not necessarily trigger an instability.  The ultimate end state of
these configurations, however, remains an open question.

Here, we explore the stability of a class of spherically symmetric
ultracompact BSs with LRs characterized by \textit{thin-shell}
profiles \cite{Collodel:2022jly} and reaching compactness $C\sim
0.32$.  We combine a set of numerical simulations in spherical
symmetry (1+1), axisymmetry (2+1), and full 3+1 dimensions with
perturbative calculations of the radial oscillation frequency,
observing excellent agreement between all approaches compatible
with the standard result that stable and unstable branches of stars,
including UCOs with LRs, are separated by local extrema in the
mass-radius diagram.  In particular, we observe no signs of instability
on the stable branch containing BH mimickers, in agreement with the
restabilization, i.e.~real frequency, of the fundamental radial
oscillation mode predicted in this regime by our perturbative
calculations.
\begin{figure}[t]
    \includegraphics[width=\linewidth]{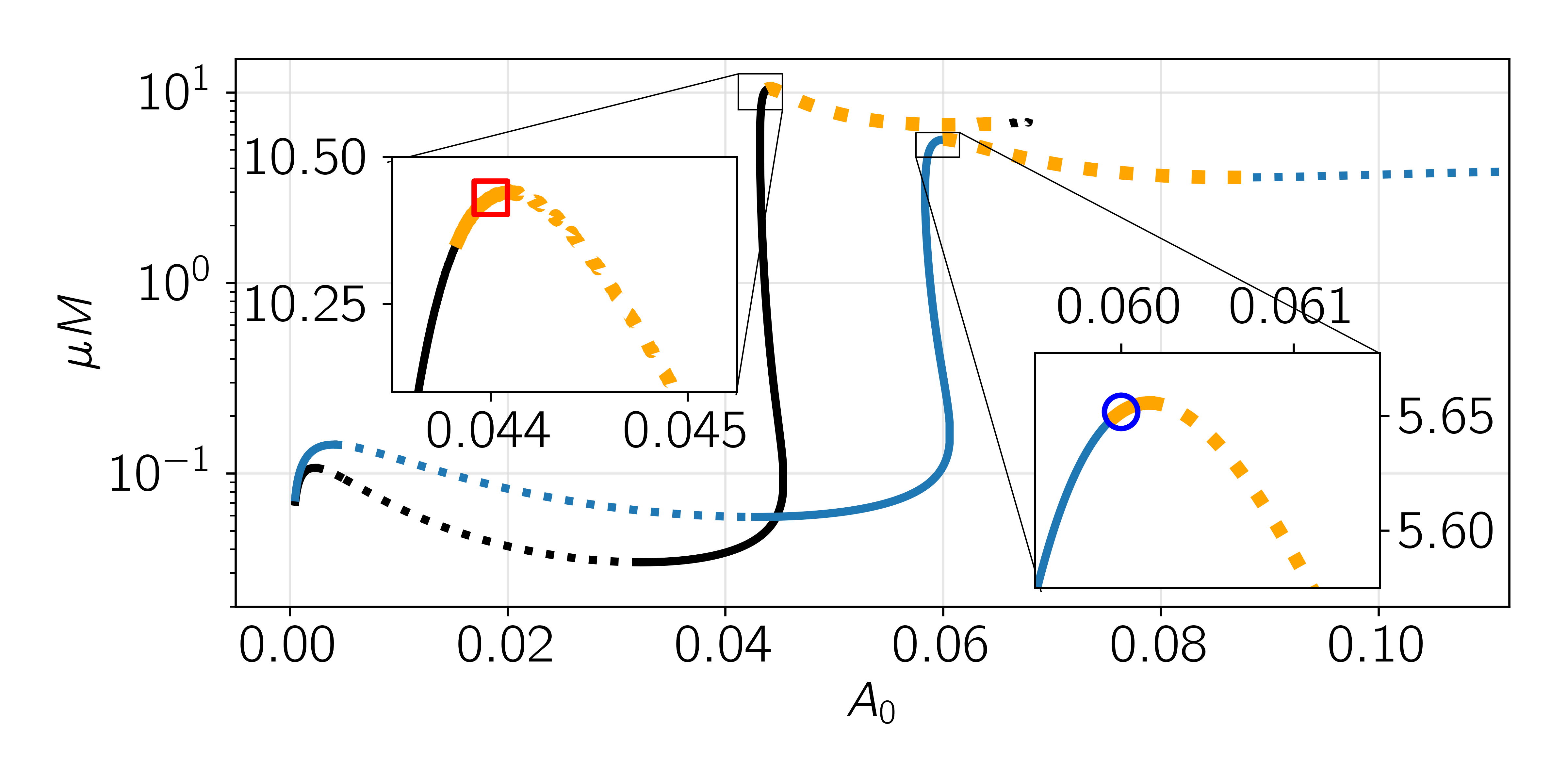}
    \caption{
    The ADM mass is shown versus the central scalar-field amplitude
    $A_0$ for the one-parameter families of BS models with solitonic
    potential parameter $\sigma_0=0.06$ (black) and $\sigma_0=0.08$
    (blue). For each family, solid (dashed) segments mark perturbatively
    stable (unstable) branches and we highlight in orange color BSs
    exhibiting a pair of LRs. The insets indicate the two perturbatively
    stable BH mimickers with parameters detailed in Table
    \ref{tab:specimen} that
    form the focus of our numerical investigation.
    }
    \label{fig:MofA}
\end{figure}
Henceforth, we employ Planck units $c=G=\hbar=1$ but keep the
scalar-field mass parameter $\mu$ which controls the scale invariance
of the spacetimes in question. Code units are translated into
SI (or similar) by fixing $\mu$ and scaling with appropriate powers
thereof: e.g.~$\mu=1.33\times 10^{-10}\,{\rm eV}$ corresponds to
$\mu^{-1}=M_{\odot}=1.48\,{\rm km}$.


{\textit{\textbf{Theory and numerical codes}}---}We consider BSs in GR
composed of one minimally coupled, complex scalar field $\varphi$,
with solitonic potential, described by the action
\begin{eqnarray}
  &&S=\int  \sqrt{-g} \left\{
  \frac{R}{16\pi}-\frac{1}{2}\left[
  g^{\mu\nu}\nabla_{\mu}\bar{\varphi}\,\nabla_{\nu}\varphi
  +V(\varphi)
  \right]
  \right\}\du^4 x\,,
  \nonumber
  \\[5pt]
  &&V(\varphi) = \mu^2 |\varphi|^2 \left(
  1-2\frac{|\varphi|^2}{\sigma_0^2}
  \right)^2\,,
  ~~~\sigma_0=\mathrm{const}\,.
  \label{eq:action}
\end{eqnarray}
Varying $S$ results in the Einstein-Klein-Gordon equations
\begin{eqnarray}
  && G_{\alpha\beta}=8\pi \,T_{\alpha\beta}\,,
  ~~~~~
  \nabla^{\mu}\nabla_{\mu}\varphi = \frac{\du V}{\du \bar{\varphi}}\,,
  \\[5pt]
  && T_{\mu\nu}
  =
  \frac{1}{2} \nabla_{(\mu} \bar{\varphi}\,\nabla_{\nu)}\varphi
  -\frac{g_{\mu\nu}}{2}
  \left[
  g^{\alpha\beta}\nabla_{\alpha}\bar{\varphi}\,\nabla_{\beta}\varphi
  + V(\varphi)
  \right].
  \nonumber
\end{eqnarray}
We compute stationary BS models in spherical symmetry writing
$\varphi(t,r)=A(r)e^{\iu \omega t}$ and using a GR version with
quadruple precision of the two-way shooting code of
Ref.~\cite{Evstafyeva:2023kfg}. The resulting single-BS spacetimes
are evolved in time using four different codes in spherical symmetry,
axisymmetry, or full 3D using the Baumgarte-Shapiro-Shibata-Nakamura
(BSSN) \cite{Baumgarte:1998te,Shibata:1995we} and the conformal
covariant $Z4$ (CCZ4) \cite{Alic:2011gg} formulations. (i) {\sc
sphericalbsevolver} achieves spherically symmetric time evolutions
through dimensional reduction with $SO(3)$ symmetry according to
Ref.~\cite{Cook:2016soy} using a uniform radial grid. (ii) A 2D
code implements axisymmetry using the same modified cartoon method
but for $SO(2)$ isometry. It is based on (iii) {\sc
ExoZvezda}~\cite{newexozvezda}, a code for exotic matter built on {\sc
GRChombo}~\cite{Radia:2021smk,Andrade:2021rbd}, which evolves 3D
domains with adaptive mesh refinement provided by {\sc chombo}
\cite{chombo}.  (iv) {\sc lean} \cite{Sperhake:2006cy} is based on
the {\sc cactus} computational toolkit \cite{Allen:1999} and evolves
3D domains with mesh refinement provided by {\sc Carpet}
\cite{Schnetter:2003rb}.  We will refer to these
codes and their respective simulations with the acronyms \texttt{Sph},
\texttt{Axi}, \texttt{E3D} and \texttt{L3D}.  Unless specified
otherwise, our simulations employ $4^{\rm th}$-order discretization, a resolution $h=1/(12\mu)$ with
outer boundary at $256\,\mu^{-1}$ on a unigrid (\texttt{Sph}) or
using 5 refinement levels (\texttt{Axi}, \texttt{E3D}, \texttt{L3D}).
\begin{table}[t]
    \centering
    \caption{
    Nickname, solitonic parameter $\sigma_0$, central
    scalar-field amplitude $A_0$, ADM mass $M$, Noether charge
    $\mathcal{N}$, the radius $r_{99}$ encompassing $99\,\%$ of the
    total mass, compactness $C\defeq M/r_{99}$ and fundamental radial oscillation
    frequency $\chi_0^2$ for the two BS models marked by square and
    circle in Fig.~\ref{fig:MofA}.
    }
    \begin{tabular}{|l|c|c|c|c|c|c|c|}
        \hline \b
        Model & $\sigma_0$ & $A_0$ & $\mu M$ & $\mu^2 \mathcal{N}$ & $\mu r_{99}$ & $C$ & $\chi_0^2/\mu^2$ \\ \hline
        \texttt{S06A044} & 0.06 & 0.044 & 10.43 & 65.9 & 31.7 & 0.33 & $2.68 \times 10^{-5}$\\ 
        \texttt{S08A06} & 0.08 & 0.06 & 5.65 & 26.1 & 17.9 & 0.31 & $7.52\times 10^{-5}$ \\  \hline
    \end{tabular}
    \label{tab:specimen}
\end{table}
%


%
\begin{figure}
    \includegraphics[width=\linewidth]{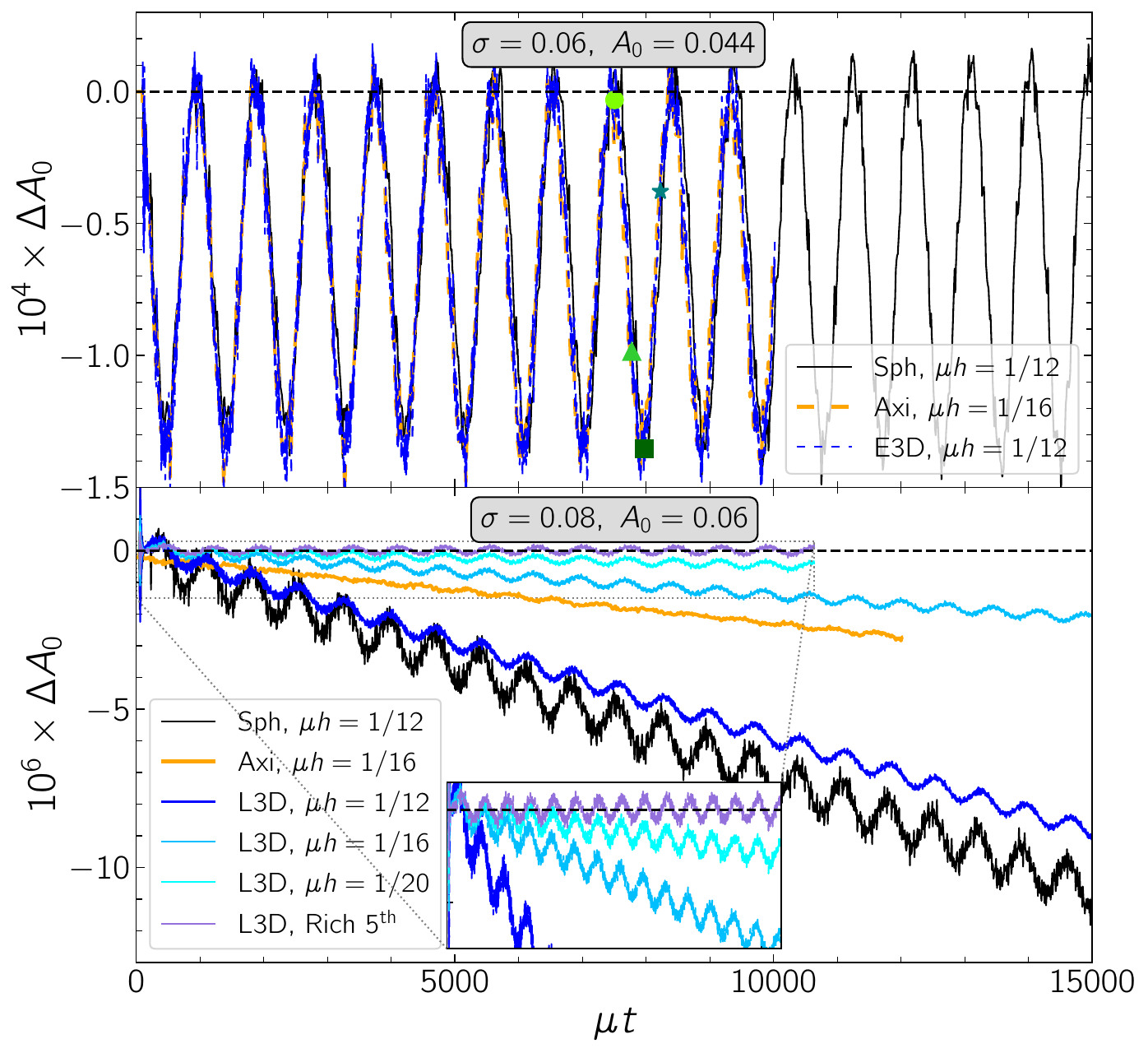}
    \caption{
    Time evolution of the deviation $\Delta A_0$ of the central
    scalar-field amplitude for the \texttt{S06A044} (top) and
    \texttt{S08A06} (bottom) models obtained in spherical symmetry,
    axisymmetry and 3D. The linear drift visible for \texttt{S08A06}
    converges away as shown in the inset.  The four symbols in the
    top panel mark times when we evaluate the LR potential;
    cf.~Fig.~\ref{fig:LR_s06_combined}.
    }
    \label{fig:123D}
\end{figure}
{\textit{\textbf{Time evolutions}---}}Our study focuses on BS families for
two values of the solitonic parameter, $\sigma_0=0.08$ and
$\sigma_0=0.06$; the resulting BS models are displayed in the plane
spanned by the central scalar amplitude $A_0$ and the Arnowitt-Deser-Misner
(ADM) mass $M$ in Fig.~\ref{fig:MofA}. There, we mark in orange
color the ultracompact BSs possessing a pair of LRs; for both
$\sigma_0$ values, this set of UCOs extends to the left, beyond the
global maximum of the $M(A_0)$ curve, i.e.~into the radially stable
branch. This regime comprises the stars of interest in our study
and we mark for each potential one representative specimen with
properties as summarized in Table \ref{tab:specimen}.

We begin our assessment of these two models by plotting in
Fig.~\ref{fig:123D} the deviation $\Delta A_0$ of the central
scalar-field amplitude from its initial value over time as obtained
with different symmetry assumptions. For the \texttt{S06A044} model,
we observe periodic oscillations with a relative magnitude $\sim\!\!\!
0.3\,\%$ and good agreement by all codes. For \texttt{S08A06}, the
oscillations are 2 orders of magnitude smaller and exhibit a linear
drift which converges away with increasing resolution.  This is
demonstrated by the three blue-shaded curves obtained with \texttt{L3D}
using different grid spacing $h$. Closer analysis of the differences
yields $5^{\rm th}$-order convergence and the Richardson extrapolated
function oscillates with high precision around zero.

Two features manifest in this figure require further explanation.
First, we attribute the considerably larger oscillations observed
for \texttt{S06A044} to a larger numerical error in the initial
data; for this model, round-off uncertainty in the BS frequency
$\sim\!\! 10^{-32}$ can effect relative variations in the ADM mass
or central lapse comparable to the level $\Delta A_0/A_0$ observed.
The \texttt{S08A06} model is less challenging and results in initial
data with small uncertainty compared to those incurred in the time
evolutions. This feature also explains the level of agreement between
time evolutions employing different symmetry: the perturbations of
\texttt{S06A044} are initial-data dominated with comparatively
smaller contributions from the time evolution.  For \texttt{S08A06},
on the other hand, details of the time evolution dominate, resulting
in more pronounced relative -- albeit tiny absolute -- differences.
This observation is confirmed by artificially lowering the quality
of the initial data or time evolution which amplifies drift or
oscillation magnitude but with no change in frequency.  The 3D
simulations displayed here have been computed using octant symmetry.
We have performed additional versions of the 3D simulations of
\texttt{S08A06} using \texttt{L3D} (\texttt{E3D}) with $h=1/12$ in
full grid mode up to $t\gtrsim 3000$ ($t\gtrsim 1000$); these give
identical results to the octant runs within round-off precision.

The time evolution of the maximal Kretschmann scalar
$R^{\alpha\beta\gamma\delta}R_{\alpha\beta\gamma\delta}$ exhibits,
up to an overall scaling, nearly identical behaviour as $\Delta
A_0$, including the oscillatory pattern, drift and $5^{\rm th}$-order
convergence toward the equilibrium value.  The Noether charge is
conserved with identical convergence to within $\Delta
\mathcal{N}/\mathcal{N} \lesssim 10^{-3}$ and $10^{-4}$ for
\texttt{S06A044} and \texttt{S08A06}, respectively, predicting
$\Delta \mathcal{N}/\mathcal{N} \lesssim 10^{-6}$ in the continuum
limit for \texttt{S08A06}.  The constraints exhibit a rapid drop
to $L_2$ norms $\sim 10^{-9}$ due to the CCZ4 damping \cite{Alic:2013xsa},
but converge only between $1^{\rm st}$ and $4^{\rm th}$ order;
cf.~Fig.~\ref{fig:2d-dtphi} in the Supplemental Material.
%
\begin{figure}[t]
  \includegraphics[width=0.48\textwidth]{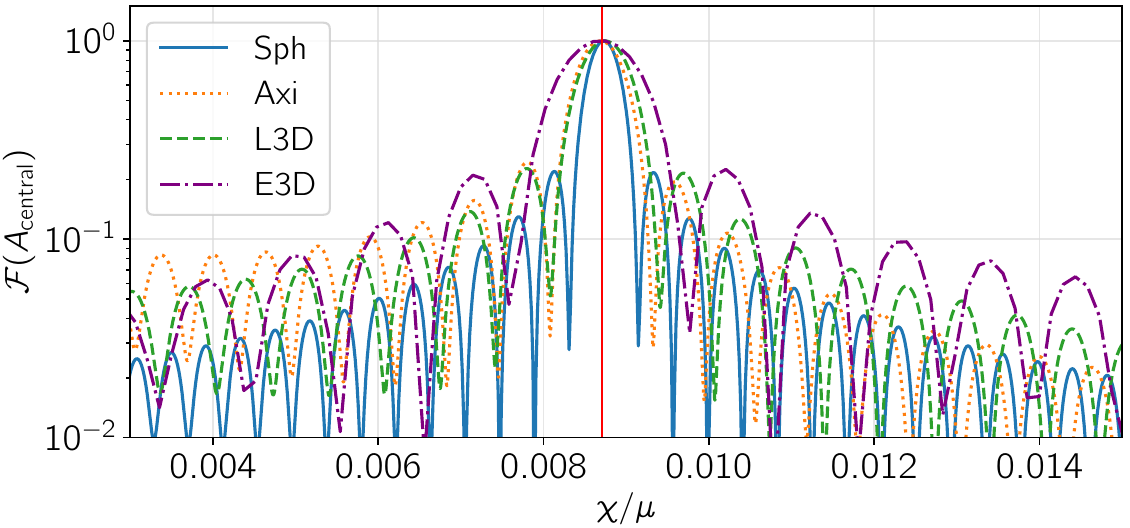}
  \caption{Power spectrum of $\Delta A_0(t)$ for \texttt{S08A060};
  the red line marks the fundamental radial mode frequency $\chi_0$.
  Units on the vertical axis are arbitrary.
  }
  \label{fig:s008_frequencies}
\end{figure}

{\textit{\textbf{Perturbative predictions}}---}The oscillatory behaviour
of $\Delta A_0$ in Fig.~\ref{fig:123D} can be understood in terms
of linearized calculations. For this purpose, we compute the frequency
$\chi_0$ of the fundamental radial oscillation mode for our BS
families using the approach of Refs.~\cite{Gleiser:1988ih,Kain:2021rmk};
cf.~Sec.~\ref{sup:Radial} in the Supplemental Material.
The transitions between stable
and unstable branches in Fig.~\ref{fig:MofA} are characterized by
zero crossings of $\chi_0^2$ such that a real (imaginary) $\chi_0$
implies radial (in)stability. For our BS models \texttt{S06A044}
and \texttt{S08A06}, we obtain $\chi_0^2>0$ as expected from their
location on the rising slope of the $M(A_0)$ curve.  The behaviour
of these BSs, as shown in Fig.~\ref{fig:123D}, exhibits excellent
agreement with the perturbative predictions: numerical imperfections
trigger perturbations around the equilibrium solution whose time
evolution is dominated by radial oscillations with the expected
frequency. We quantify this agreement by plotting in
Fig.~\ref{fig:s008_frequencies} for \texttt{S08A06} the Fourier
spectrum of $\Delta A_0(t)$, i.e.~the data of Fig.~\ref{fig:123D}.
The spectrum peaks at the $\chi_0$ value predicted by the perturbative
calculation.  The corresponding analysis for \texttt{S06A044} yields
similar agreement; cf.~Fig.~\ref{fig:s006_frequencies} in
the Supplemental Material.
%
\begin{figure}[t]
    \centering
    \includegraphics[width=\linewidth]{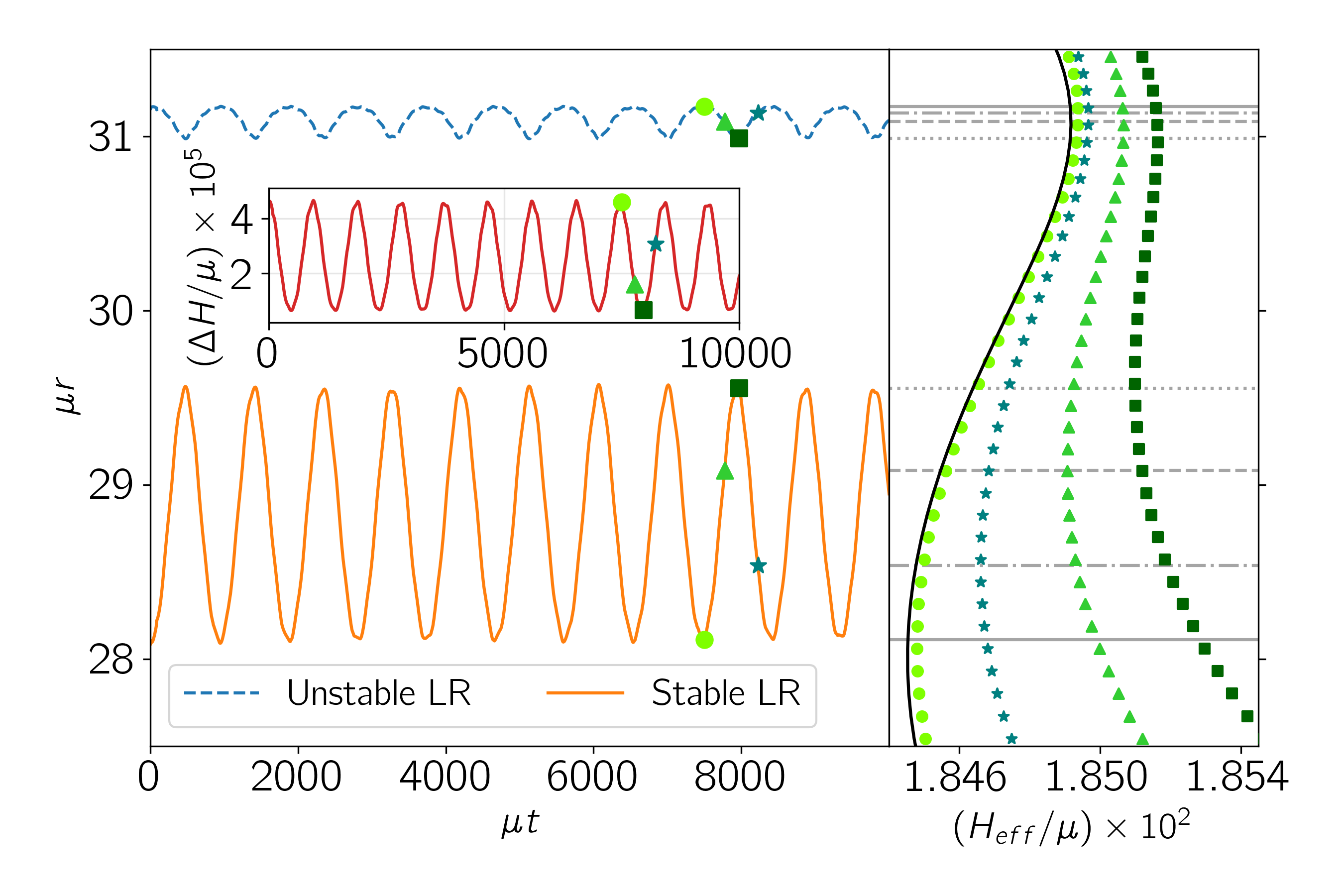}
    \caption{\textit{(left)} Areal radius of the (un)stable light
    ring over time as extracted from the AEP $H_{\rm eff}$
    \textit{(right)} during the time evolution of \texttt{S06A044}
    performed with \texttt{E3D}.  The inset shows the difference
    between the extrema of the AEP over time. The green markers
    correspond to the snapshots of the AEP given in the right panel.
    The gray lines indicate the positions of the extrema. The solid
    line is the effective potential \eqref{eq: eff pot} at the start
    of the simulation.
    }
    \label{fig:LR_s06_combined}
\end{figure}

{\textit{\textbf{Light-ring structure}}---}Having demonstrated the long-term
stable behaviour of our BH mimickers \texttt{S06A044} and
\texttt{S08A06}, we now turn to the question whether their LR
structure remains preserved.  In a stationary and axisymmetric
spacetime, a LR is defined as a null geodesic with a tangent vector
field that is always a linear combination of the Killing vectors
$\partial_t$ and $\partial_{\phi}$, assuming coordinates
$(t,r,\theta,\phi)$ adapted to the symmetry. The conditions for a
LR can then be recast as follows [cf.~Eqs.~(1)-(8) in Ref.~\cite{Cunha:2017qtt}]:
%
\begin{equation}\label{eq: eff pot}
  \partial_{\mu}H_{\pm}=0\,,
  ~~\text{where}
  ~~
  H_{\pm}
  \defeq
  \frac{-g_{t\phi}\pm\sqrt{g_{t\phi}^2-g_{tt}g_{\phi\phi}}}{g_{\phi\phi}}\,.
\end{equation}
Furthermore, (un)stable LRs are located at local (maxima) minima
of $H_{\pm}$.  In spherical symmetry with adapted coordinates,
$H_{\pm}$ simplifies to a single potential function
$H=\sqrt{-g_{tt}/g_{\phi\phi}}$.

Under the assumption that an evolved spacetime remains nearly
spherically symmetric and stationary, the potential function can
be approximated by an \textit{adiabatic effective potential} (AEP)
$H_{\rm eff}$ computed via an averaging procedure analogous to
Ref.~\cite{Cunha:2022gde}.  Specifically, we transform between
standard Cartesian and spherical coordinates $x^{\alpha}=(t,x,y,z)$,
$\hat{x}^{\mu}=(t,R,\theta, \phi)$. For a sphere of fixed coordinate
radius $R$, we compute the proper area $\mathcal{A}$, the areal
radius $r\defeq \sqrt{\mathcal{A}/(4\pi)}$, and the AEP $H_{\rm
eff}(r) = \sqrt{(\langle\alpha^2 \rangle-\langle \gamma_{ij}
\beta^i\beta^j\rangle)}/r$, where $\alpha$, $\beta^i$ and $\gamma_{ij}$
are the lapse, shift and spatial metric in Cartesian coordinates,
and $\langle\,.\,\rangle$ denotes averaging over the sphere. Repeating
this procedure for a sequence of $\sim 50$ coordinate spheres with
$\Delta R\approx 0.1$ across the relevant shell region, we obtain
at each time $t$ an estimate for the potential $H_{\rm eff}(r)$.
We validate the robustness of this method and the assumptions in
Sec.~\ref{sup:Adiabatic} of the Supplemental Material.

Our analysis of the resulting potentials for \texttt{S06A044} and
\texttt{S08A06} indeed yields one minimum and one maximum at all
$t$. The locations of the respective LRs exhibit minor periodic
variation in line with the BS oscillations observed in the
other diagnostics, but the pair of LRs remains in place
with no indication of instability.  We show these variations together
with selected profiles of the AEP over a representative oscillation
cycle in Fig.~\ref{fig:LR_s06_combined} for \texttt{S06A044}.  The
same behaviour, merely with smaller oscillation amplitude, is found
for \texttt{S08A06}.

%
\begin{figure}[t]
    \centering
    \includegraphics[width=\linewidth]{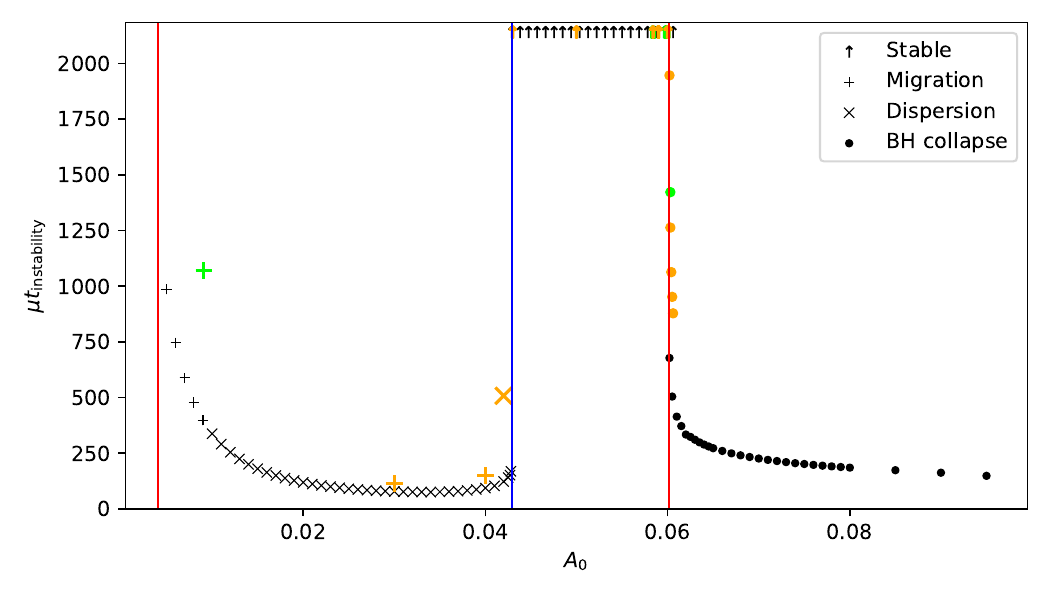}
    \caption{Instability timescales as a function of $A_0$ for
    solitonic BSs with $\sigma_0=0.08$.  The vertical lines mark
    the transition between stable and unstable branches as predicted
    by linear theory; cf.~Fig.~\ref{fig:MofA}. The marker style
    indicates dynamical fate while the color indicates the code
    used: black for \texttt{Sph}, lime green for \texttt{Axi},
    orange for \texttt{L3D}. BS models whose time evolution show
    no sign of instability are represented with arrows at the top
    of the figure.
    }
    \label{fig:migration_times}
\end{figure}

{\textit{\textbf{Nonlinear stability}}---}We complement our detailed analysis
of the BS models \texttt{S06A044} and \texttt{S08A06} with a larger
set of time evolutions of the $\sigma=0.08$ BS sequence. We have
used all codes for this analysis, but restricted 3D evolutions to
$\mu t\lesssim 3000$ for reasons of computational cost. In this
analysis, we define the instability time as the moment when $A_0(t)$
departs by $10\,\%$ from its equilibrium value.  If $A_0(t)$ yields
no signature of departing from the oscillatory pattern seen in
Fig.~\ref{fig:123D} within $3000\,\mu^{-1}$, the model is designated
stable. The result is shown in Fig.~\ref{fig:migration_times} and
fully agrees with the perturbative predictions: models on unstable
branches succumb to exponentially growing perturbations on a timescale
that diverges at the threshold to stable branches.  These
transition points coincide with extrema in the $M(A_0)$ curves,
i.e.~points where the oscillation frequency $\chi_0$ of the fundamental
mode switches between real and imaginary.

\newpage
{\textit{\textbf{Concluding remarks}}---}We have performed a series of fully
nonlinear evolutions of the Einstein-Klein-Gordon equations governing
the dynamics of solitonic boson stars in the thin-shell regime in
3+1 dimensions imposing spherical, axi-, octant or no symmetry.  A
subset of the models under consideration are ultracompact, exhibiting
a pair of light rings, one of which is stable. We have evolved these
objects for $\gtrsim$1000 light crossing times, finding no evidence
of an instability due to the appearance of the stable light ring,
as previously conjectured \cite{Keir:2014oka, Cardoso:2014sna} and
reported for spinning BSs \cite{Cunha:2022gde}.  We observe excellent
agreement between the axisymmetric/3D evolutions and those in
spherical symmetry, suggesting that the inclusion of nonspherical
degrees of freedom does not significantly affect the stability of
these models. Taking into account also the agreement of our numerical
results with perturbative predictions for stability and fundamental
radial oscillation frequencies, we interpret our findings as strong
numerical evidence in favor of the stability of this class of BH
mimickers.

As a cautionary note, we add that poor choices of numerical
ingredients, in particular the constraint-damping parameter $\kappa_1$
of the CCZ4 formulation, can result in seemingly plausible instabilities
of our thin-shell BSs. With $M\kappa_1 \gg 0.1$, for example, both
models \texttt{S06A044} and \texttt{S08A06} collapse to a BH; the
erroneous nature of these simulations reveals itself only through
large and growing constraint violations and complete lack of
convergence.

\begin{figure}[H]
    \centering
    \includegraphics[width=0.96\linewidth]{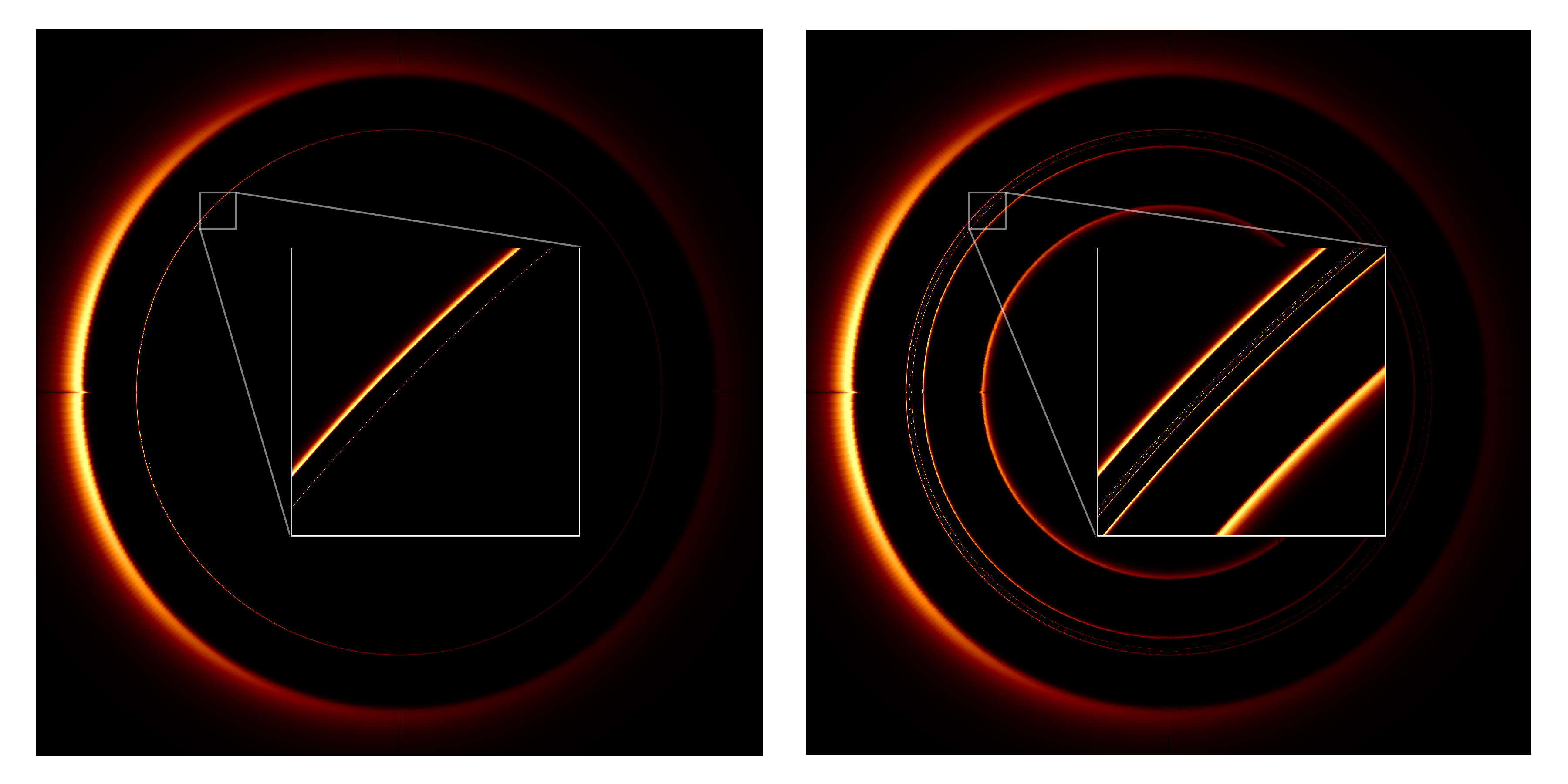}
    \caption{Side-by-side comparison of ray-traced images of
    (\textit{left}) a Schwarzschild BH and (\textit{right)} the
    \texttt{S06\_A044} model of the same ADM mass. The color scale
    indicates the observed brightness intensity, in arbitrary units,
    from a simple geometrically thin disk located in the equatorial
    plane. Both panels also provide a close-up of the photon ring
    region, exhibiting their fractal structure. Images are obtained
    with \texttt{FOORT} \cite{newMayerson:2024fgh, Mayerson:2025foo}.}
    \label{fig: ray trace}
\end{figure}

If the presence of the stable light ring is not sufficient to destroy
these solitonic boson stars, this family of spherically symmetric
models, i.e.~with quantized spin zero, as well as similar UCOs, can
be considered as viable black-hole mimickers again.  The thin-shell
solutions under consideration here can mimic the phenomenology of
black holes in several different ways, as for example in the GW
emission from binaries \cite{Evstafyeva:2024qvp} or the creation
of BH like shadows in VLBI images \cite{Rosa:2023qcv}. Figure
\ref{fig: ray trace} illustrates how the \texttt{S06A044} model
could resemble a Schwarzschild BH with the same ADM mass in the
presence of an equatorial, optically thin accretion disk. The image
of the BS clearly displays additional structure toward the center
of the image, which is a result of the absence of a horizon as well
as the pair of light rings. The outermost direct emission band, as
well as the first lensed photon ring are nearly indistinguishable
for both images.

{\textit{\textbf{Acknowledgments}}---}G.A.M. is supported by the Cambridge
Trust at the University of Cambridge. S.S. is supported by the
Centre for Doctoral Training at the University of Cambridge funded
through STFC. T.E. acknowledges the Perimeter Institute for Theoretical
Physics, supported by the Government of Canada through the Department
of Innovation, Science and Economic Development and by the Province
of Ontario through the Ministry of Colleges and Universities. We are grateful to William Janczewski for pointing out an error in an earlier version of this manuscript. This
work has been supported by STFC Research Grant No. ST/V005669/1.
We acknowledge support by the NSF Grants No.~PHY-090003,~No.~PHY-1626190,
and No.~PHY-2110594; DiRAC projects No.~ACTP284 and No.~ACTP238; STFC capital
Grants No.~ST/P002307/1, No.~ST/R002452/1, No.~ST/I006285/1, and
No.~ST/V005618/1; and STFC operations Grant No.~ST/R00689X/1. Computations
were done on the CSD3 and Fawcett (Cambridge), Cosma (Durham),
Niagara (Toronto), Narval (Montreal), Stampede2 (TACC) and Expanse
(SDSC) clusters.

%


\onecolumngrid
\newpage
\begin{center}
  \textbf{\large{Supplemental Material}} \\
\end{center}
\twocolumngrid

\setcounter{equation}{0}
\setcounter{figure}{0}
\setcounter{table}{0}
\setcounter{page}{1}
\makeatletter
\renewcommand{\theequation}{S\arabic{equation}}
\renewcommand{\thefigure}{S\arabic{figure}}
\renewcommand{\bibnumfmt}[1]{[S#1]}
\renewcommand{\citenumfont}[1]{S#1}

\section{Convergence Analysis}

\begin{figure}[b]
\centering
  \includegraphics[width=0.45\textwidth]{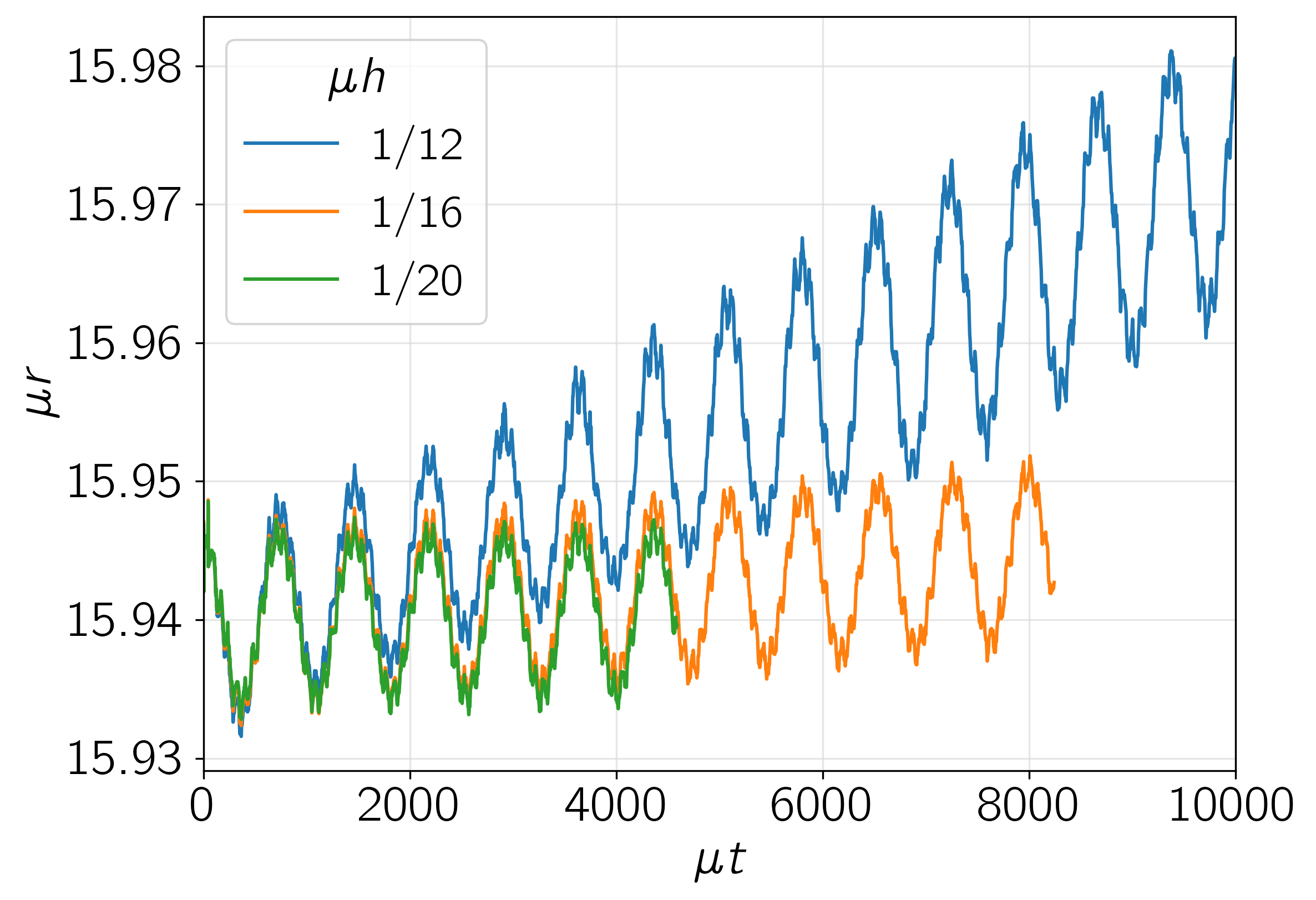} 
  \caption{Evolution of the areal radius of the stable LR for the
  \texttt{S08A06} model obtained with the \texttt{E3D} code using
  different resolutions indicated in the legend. The differences
  between the curves yield convergence at 5\textsuperscript{th}
  order.}
  \label{fig: s08 LR convergence}
\end{figure}

\textit{Light rings:} We check convergence of the computed light
ring (LR) radii for the \texttt{S08A06} model by comparing results
obtained at resolutions $\mu h \in \{1/12, 1/16, 1/20\}$ and display
in Fig.~\ref{fig: s08 LR convergence} the resulting areal radii of
the stable LR as functions of time. At low resolution the LR radius
exhibits a linear drift, much like what was observed for the central
amplitude and Kretschmann scalar (cf.~Fig.~2 in the main text).
Again, this drift converges away at 5\textsuperscript{th} order,
and the Richardson extrapolated values, obtained from the two
highest-resolution runs and displayed in Fig.~\ref{fig: s08 LR
Richardson}, oscillate around constant values.
\begin{figure}[t]
\centering
  \includegraphics[width=0.45\textwidth]{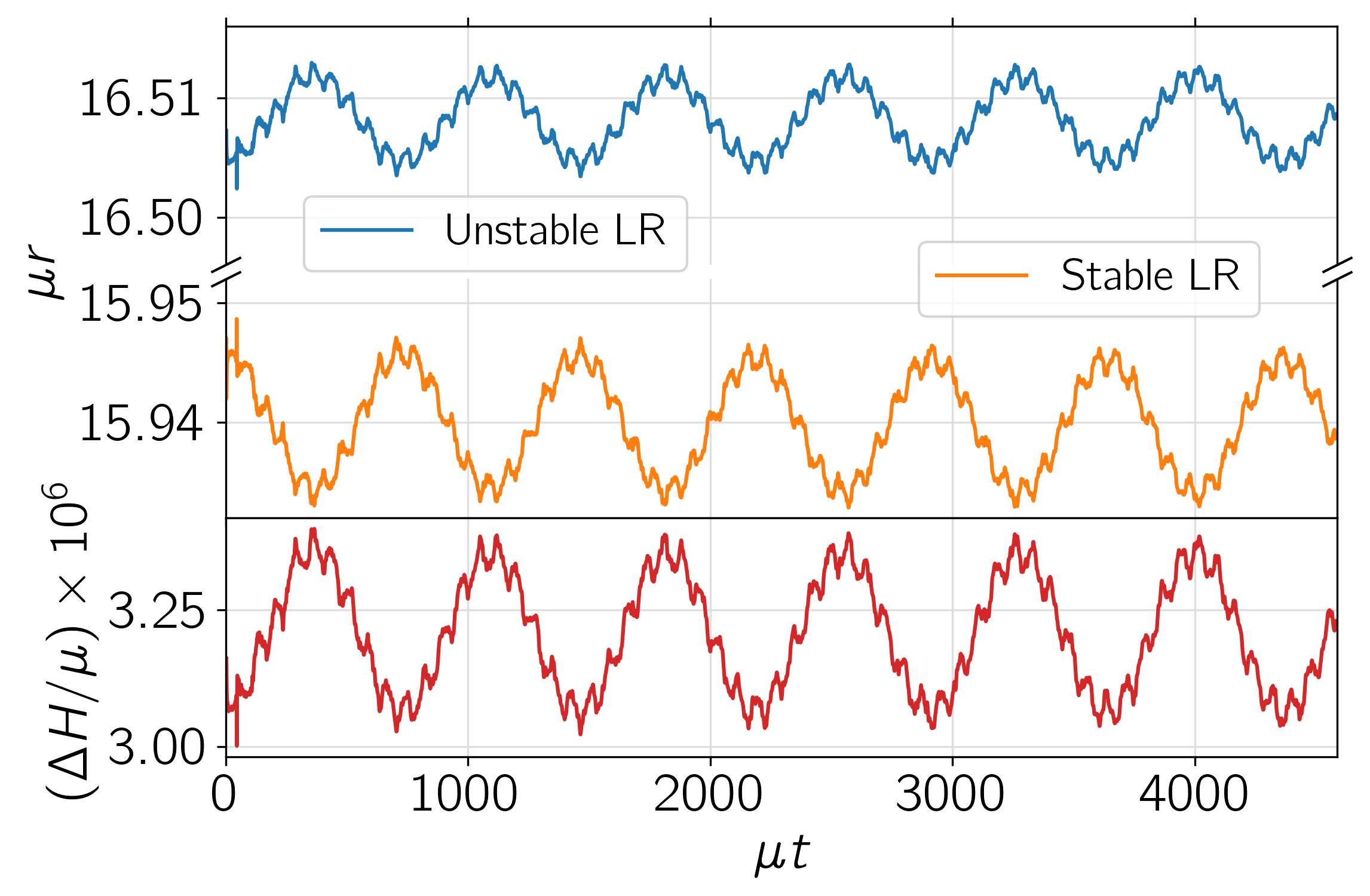} 
  \caption{Richardson extrapolation (5\textsuperscript{th} order)
  of the LR areal radii of the \texttt{S08A06} model based on the
  runs at resolutions $\mu h = 1/16, 1/20$. The bottom panel shows
  the Richardson extrapolated evolution of the potential difference
  between the extrema of the AEP.}
  \label{fig: s08 LR Richardson}
\end{figure}

For the \texttt{S06A044} model, we observe no discernible drift and
increasing the resolution merely results in convergence of the
regular oscillations. Again, this underlines that the oscillatory
perturbation of the \texttt{S06A044} model is dominated by the
quality of the initial data rather than details of the evolution.

In Fig.~\ref{fig: s08 LR Richardson} we see that the oscillation
amplitude of the stable LR is slightly larger than that of the
unstable LR. This effect is more pronounced for the \texttt{S06A044}
model displayed in Fig.~4 of the main text. The reason in both cases
is that the unstable light ring is - to good approximation - outside
the thin shell, fixing its location close to the Schwarzschild value
by Birkhoff's theorem. The stable light ring, on the other hand,
is located in the interior of the star, and therefore affected more
strongly by the perturbation.
\begin{figure*}[t]
    \centering
    \includegraphics[width=1\linewidth]{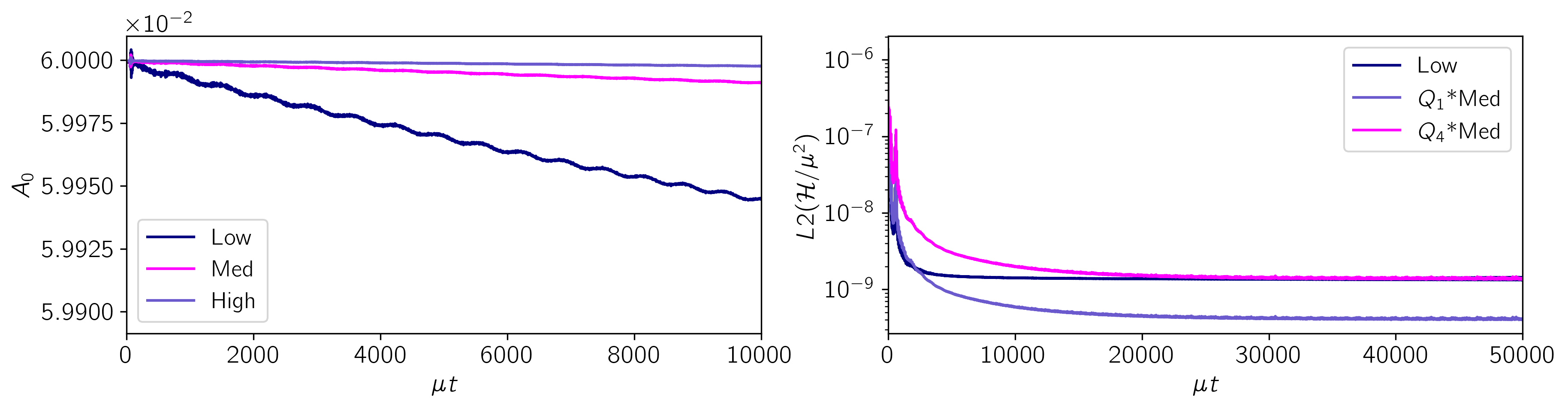}\\
    \includegraphics[width=1\linewidth]{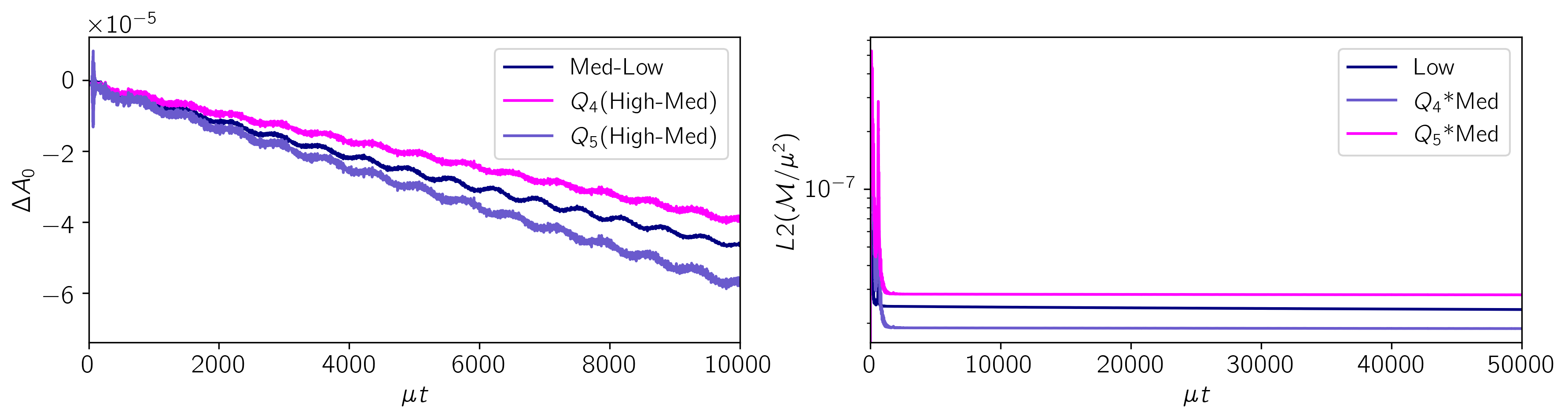}
    \caption{The left (right) panels illustrate convergence results
    obtained from \texttt{Axi} simulations of the \texttt{S08A06}
    model for the scalar field amplitude (volume-weighted $L_2$
    norms of the Hamiltonian and momentum constraints).  We amplify
    the differences and constraint magnitudes by factors $Q_n$ as
    expected for $n^{\rm th}$-order convergence.  }
    \label{fig:2d-dtphi}
\end{figure*}

\textit{2D code}: We next discuss the convergence results for the
\texttt{S08A06} model obtained in axisymmetry with the \texttt{Axi}
code. Using $\mu h= 1/8$, $\mu h = 1/12$ and $\mu h = 1/16$ on the
innermost refinement level for the low, medium and high resolutions,
respectively, we obtain convergence between fourth and fifth order
for the scalar field amplitude; cf.~top panel of Fig.~\ref{fig:2d-dtphi}.
In the bottom panels of the figure, we also monitor the volume-weighted
$L_2$ norms of the constraint violations. We find first-order
convergence in the Hamiltonian constraint at very early times which
gradually transitions to fourth order at $\mu t \approx 20000$ where
it persists throughout the remainder of the simulation. The momentum
constraint $\mathcal{M} \defeq (\mathcal{M}_x^2+\mathcal{M}_y^2)^{1/2}$
on the other hand rapidly reaches convergence between fourth and
fifth order.

\section{Additional results for $\sigma_0 = 0.06$}

\begin{figure}
  \includegraphics[width=0.48\textwidth]{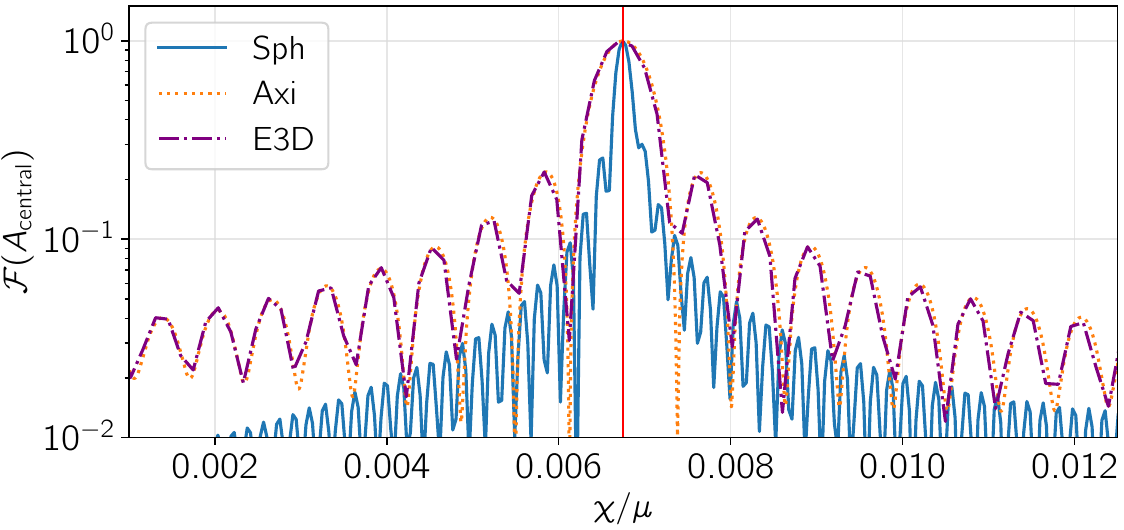}
  \caption{Power spectrum of $\Delta A_0(t)$ for the BS model
  \texttt{S06A044}; the red line marks the fundamental
  radial mode frequency $\chi_0$. Units on the vertical axis are arbitrary.
  }
  \label{fig:s006_frequencies}
\end{figure}

In this section we present results for the $\sigma_0 = 0.06$ family
of models, supplementing those presented for $\sigma_0 = 0.08$ in
the main text. In Fig.~\ref{fig:s006_frequencies}, we compare the
power spectrum of central scalar-amplitude oscillations to the
radial oscillation frequency, computed with perturbation theory.
In this case the larger error in our initial data causes the system
to migrate early on to a boson-star (BS)  model with slightly smaller
central scalar amplitude than the initial value $A_0=0.044$. Following
Ref.~\cite{SMKain:2021rmk} we take this effect into account by computing
the radial frequency for a BS model with an amplitude averaged over
several oscillation cycles, $\langle A_0 \rangle = 0.04394$.  Once
again, we find very good agreement between results from the
perturbative analysis and the codes enforcing different symmetry
assumptions.

\begin{figure}[t]
    \centering
    \includegraphics[width=\linewidth]{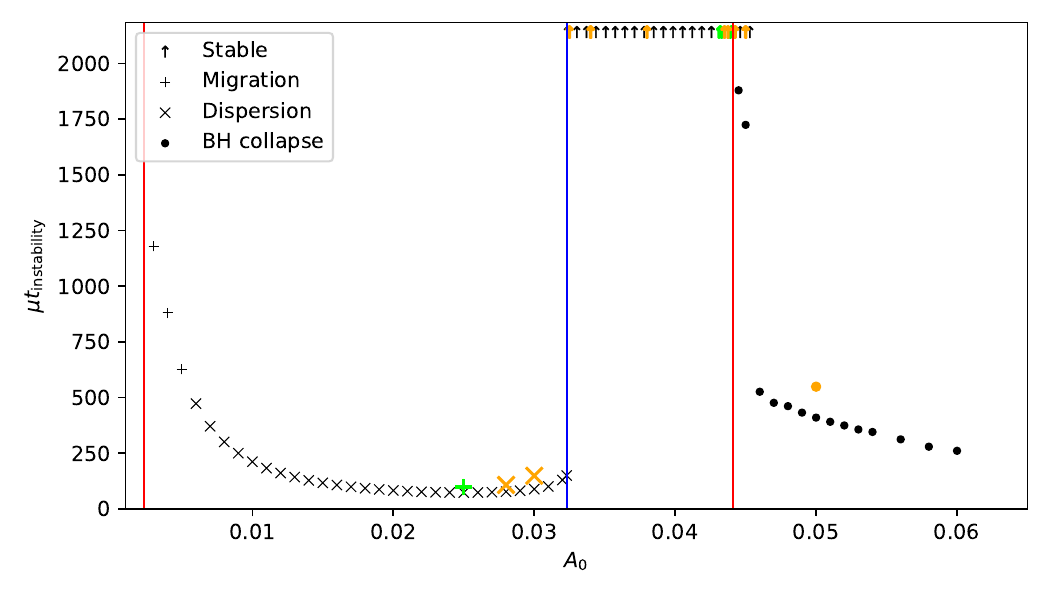}
    \caption{Instability timescales as a function of $A_0$ for
    solitonic BSs with $\sigma_0=0.06.$ The vertical lines mark the
    transition between stable and unstable branches as predicted
    by linear theory. The marker style indicates the dynamical fate
    while colour indicates the code used: black for \texttt{Sph},
    lime green for \texttt{Axi}, orange for \texttt{E3D}.  BS models
    whose time evolution shows no sign of instability are represented
    with arrows at the top of the figure.}
    \label{fig:migration_times_s006}
\end{figure}

In Fig.~\ref{fig:migration_times_s006}, we show timescales of
instability for a sample of BS models with $\sigma_0 = 0.06$. As
before, we use the time at which the absolute change in the central
amplitude $|\Delta A_0(t)|$ exceeds $10 \%$ of the initial central
amplitude $\Delta A_0(0).$ Once again, we find that the predictions
of the linear radial perturbation theory are borne out by all codes
regardless of symmetry assumptions. Models on the first unstable
branch either migrate or disperse depending on the sign of the
binding energy, and models on the second unstable branch always
collapse to a BH. We emphasize that the specific time values depend
on factors such as resolution, gauge, evolution method and numerical
parameter choices or mesh refinement; values determined with different
codes therefore exhibit some variation but always agree on stability
or instability. Note also that for some values of $A_0$, multiple
BS models exist.  In those cases, the two lower-mass models are
always stable while the heaviest BS may be stable or not, in
accordance with the predictions of linear theory.

\section{Radial Oscillation Frequencies}
\label{sup:Radial}
In this section we discuss in detail the method used to compute
fundamental radial oscillation frequencies for solitonic BSs.  Here,
we use overdots to denote time derivatives and primes for radial
derivatives.

We begin with a spherically symmetric metric ansatz,
\begin{equation}\label{eq: BS metric ansatz}
    \du s^2 = -\alpha^2(t,r) \du t^2 + X^2(t,r)\du r^2 + r^2\du \Omega_2^2.
\end{equation}
Decomposing the complex scalar field into two real fields, $\varphi(t,r)
= \varphi_1 + i\varphi_2$, the Einstein-Klein-Gordon equations
become,
\begin{align}
  \frac{X'}{X} =\, & \frac{1 - X^2}{2r} + 2\pi r X^2\left(
  \frac{|\dot{\varphi}|^2}{\alpha^2} + \frac{|\varphi'|^2}{X^2} + V \right),
  \label{eq:bg_X} \\
  \frac{\alpha'}{\alpha} =\, & -\frac{1 - X^2}{2r} + 2\pi r X^2\left(
  \frac{|\dot{\varphi}|^2}{\alpha^2} + \frac{|\varphi'|^2}{X^2} - V \right)
  \label{eq:bg_alpha}, \\
  \varphi_i'' =\, & \left(\frac{X'}{X} - \frac{\alpha'}{\alpha}
  - \frac 2r  \right)\varphi_i' +
  \frac{X^2 \dot\alpha + X\alpha\dot X}{\alpha^3} \dot\varphi_i
  + \frac{X^2}{\alpha^2}\ddot\varphi_i, \label{eq:phi_eq} \\
  \alpha'' =\, & \frac{X\ddot{X}}{\alpha}
  - \frac{X\dot X \dot\alpha}{\alpha^2} + \frac{X'\alpha'}{X}
  - \frac{\alpha'}{r} + \frac{\alpha X'}{rX} \label{eq:Ethetatheta} \\
  &+ 4\pi\alpha \left( \frac{X^2}{\alpha^2}|\dot{\varphi}|^2
  - |\varphi '|^2 - 4X^2V \right) \nonumber .
\end{align}
We solve Eqs.~\eqref{eq:bg_X}-\eqref{eq:phi_eq}
using the ansatz $\varphi(t,r) = A(r) e^{i \omega t}$
to obtain a background solution  henceforth denoted by $\{A_b(r), \alpha_b(r), X_b(r), \omega\}$. The background scalar-field amplitude in particular satisfies 
\begin{equation}
  A_b'' = - \frac{\omega^2X_b^2}{\alpha_b^2}  A_b
  +  A_b' \left(\frac{X'_b}{X_b} - \frac{\alpha'_b}{\alpha_b} - \frac{2}{r}
  \right) + X_b^2 A_b V'(A_b^2)\,,\label{eq:bg_A}\,.
\end{equation}
We perturb the BS background by introducing the perturbations
$\delta \psi_1$, $\delta \psi_2$, $\delta \nu$ and $\delta \lambda$
according to
\begin{align}
  \varphi & = e^{i \omega t} \left( \psi_1 + i\psi_2 \right)\,, &
  \nonumber\\[5pt]
  \psi_1 & = A_b(1 + \delta\psi_1), \qquad & \psi_2  & =A_b \delta \psi_2\,,
  \nonumber \\
  \alpha & = \alpha_b\left(1 + \frac 12 \delta\nu \right), \qquad \  &
  X &= X_b\left(1 + \frac 12 \delta\lambda \right). \label {eq:perts}
\end{align}
Substituting this expansion into Eqs.~\eqref{eq:bg_X}, \eqref{eq:phi_eq},
\eqref{eq:Ethetatheta}, taking the difference between
Eqs.~\eqref{eq:bg_alpha} and \eqref{eq:bg_X}, and linearizing, we
obtain
\begin{widetext}
  \begin{align}
  \delta\psi_1'' =\,& X_b^2\left(1 - \frac{\omega^2}{\alpha_b^2}
  - \frac{8A_b^2}{\sigma_0^2} + \frac{12A_b^4}{\sigma_0^4}\right)\delta\lambda
  + \frac{X_b^2}{\alpha_b^2}\delta \ddot\psi_1
  + \frac{16A_b^2X_b^2}{\sigma_0^2}\left( \frac{3A_b^2}{\sigma_0^2} - 1
  \right)\delta\psi_1 + \frac{A_b'}{2A_b}\left(\delta\lambda' + 4\delta\psi_1'
  - \delta\nu' \right) \nonumber \\
  &+ \frac{\omega X_b^2}{\alpha_b^2}\left(2 \delta\dot\psi_2 - \omega\delta\nu
  \right) + \left(\frac{X_b'}{X_b} - \frac{\alpha_b'}{\alpha_b} - \frac 2 r
  \right)\delta\psi_1' \,, \label{eq:psi1_rr} \\
  \delta\lambda' =\,& \left( \frac{2X_b'}{X_b} - \frac 1r - 4\pi r (A_b')^2
  \right)\delta\lambda + 8\pi rX_b^2 \left(A_b^2 + \frac{(A_b')^2}{X_b^2}
  + \frac{\omega^2A_b^2}{\alpha_b^2} + \frac{12A_b^6}{\sigma_0^4}
  - \frac{6A_b^4}{\sigma_0^2} \right)\delta\psi_1 \nonumber\\
  &+ \frac{4\pi r\omega A_b^2 X_b^2}{\alpha_b^2}\left(\omega\delta\nu
  + 2 \delta\dot\psi_2 \right) + 8\pi r A_b'A_b\delta\psi_1'.
  \label{eq:lambda_r} \\
  \delta\ddot\lambda =\,& \frac{2\alpha_b}{X_b^2}\left[
  \frac{\alpha_bX_b'}{rX_b} + \alpha_b'\left(\frac{X_b'}{X_b}
  - \frac{1}{r}\right)- \alpha_b'' - 4\pi\alpha_b (A_b')^2 \right]
  \delta\lambda  - \frac{16\pi \alpha_b^2}{\omega X_b}\left[
  \frac{(A_b')^2}{X_b^2} + A_b^2 - \frac{8A_b^4}{\sigma_0^2}
  + \frac{12A_b^6}{\sigma_0^4} - \omega^2A_b^2\right]\delta\psi_1
  \nonumber \\
  &+ 8\pi\omega A_b^2\left(\omega\delta\nu - 2\delta\dot\psi_2\right)
  + \frac{1}{rX_b}\left(\frac{r\alpha_b^2\delta\nu'}{X_b} \right)'
  + \frac{16\pi\alpha_b^2A_b'A_b}{X_b^2}\delta\psi_1' - \frac{1}{rX_b^2}
  \left(\alpha_b^2 + r\alpha_b\alpha_b' \right)\delta\lambda'.
  \label{eq:lambda_tt} \\
  \delta\lambda' - \delta\nu' \label{eq:nu_r} =\,& 2\left(\frac{X_b'}{X_b}
  - \frac{\alpha_b'}{\alpha_b} - \frac1 r \right)+ 16\pi rA_b^2X_b^2
  \left(1 - \frac{8A_b^2}{\sigma_0^2} + \frac{12A_b^4}{\sigma_0^4}  \right).
\end{align}
\end{widetext}
Next, we make the harmonic ansatz $\delta\psi_1(t,r) = e^{\iu \chi
t}f(r),$   $\delta\lambda(t,r) = e^{\iu \chi t} g(r) $ with $\chi\in
\mathbb{C}$ to be determined. The resulting system of two pulsation
equations is self-adjoint, so that, in analogy to the BS frequencies
$\omega$, the radial oscillation frequencies $\chi$ form an infinite
discrete sequence $\chi_0^2 < \chi_1^2 < ...$, where the solution
with eigenvalue $\chi_n^2$ is regular with $n$ zero crossings in
$f$ and $g$.  A negative $\chi_0^2$ corresponds to a radial
instability, while $\chi_0^2 > 0$ guarantees (linear) radial
stability.

To obtain the first pulsation equation, we solve Eq.~\eqref{eq:lambda_r}
for $\delta\dot\psi_2$ and substitute into Eq.~\eqref{eq:psi1_rr}
to obtain an equation that depends on $\delta\nu$ only through its
radial derivative which, in turn, can be removed using Eq.~\eqref{eq:nu_r}.
For the second pulsation equation, we combine Eqs.~\eqref{eq:bg_A},
\eqref{eq:lambda_r} and \eqref{eq:lambda_tt} to obtain an equation
that depends on $\delta \nu'$ and $\delta \nu''$ but not $\delta
\nu$ directly. By using Eq.~\eqref{eq:nu_r}, its derivative and
Eq.~\eqref{eq:bg_A}, we can eliminate $\delta \nu$ completely and
thus arrive at our final system of two ODEs,
\begin{widetext}
  \begin{align}
  \label{eq:pulsation1} f'' =\, & X_b^2  \bigg[2 +\frac{2(A_b')^2}{X_b^2A_b^2}
  + \frac{2\omega^2 - \chi^2}{\alpha_b^2} + 8\pi rA_bA_b'
  - \frac{32A_b^2}{\sigma_0^2}\left(1 + 2\pi r A_b^2A_b'  \right)
  + \frac{8A_b^4}{\sigma_0^4}\left(9 + 12\pi rA_bA_b' \right) \bigg]f \\
  &+ \bigg[ X_b^2\left(1 - \frac{8A_b^2}{\sigma_0^2}
  + \frac{12A_b^4}{\sigma_0^4} \right) - \frac{1}{4\pi r^2 A_b^2}
  - \frac{(A_b')^2}{A_b^2}  - \frac{A_b'}{rA_b}
  - \frac{A_b'\alpha_b'}{A_b\alpha_b} + \frac{X_b'}{X_bA_b}\left(
  A_b' + \frac{1}{2\pi r A_b} \right) \bigg] g \nonumber \\
  &+ \left(\frac{X_b'}{X_b} - \frac{\alpha_b'}{\alpha_b} - \frac 2r \right)f'
  - \frac{1}{4\pi r A_b^2}g', \nonumber
\end{align}
\begin{align}
  g'' =\,& 32\pi r X_b^2  \bigg[A_bA_b' + \frac{\alpha_b'A_b^2}{\alpha_b}
  - \label{eq:pulsation2} \frac{8A_b^3}{\sigma_0^2}\left(2A_b'
  + \frac{A_b\alpha_b'}{\alpha_b} \right)    +  \frac{X_b'}{X_b}\left(A_b^2
  - \frac{4A_b^4}{\sigma_0^2} + \frac{6A_b^6}{\sigma_0^4}
  - \frac{(A_b')^2}{X_b^2} \right)
  \\ &
  + \frac{12A_b^5}{\sigma_0^4}\left(3A_b' + \frac{A_b\alpha_b'}{\alpha_b}
  \right)    \bigg] f  + 3\left(\frac{X_b'}{X_b} - \frac{\alpha_b'}{\alpha_b}
  \right)g '
  \nonumber
  + \bigg[16\pi (A_b')^2  - \frac{\chi^2X_b^2  - 2 (\alpha_b')^2}{\alpha_b^2}
  - \frac{4\alpha'_0}{r\alpha_b} + \frac{2}{r^2} \\ \nonumber &+\frac{2}{X_b}
  \left(X_b'' + \frac{2\alpha_b'X_b'}{\alpha_b} - \frac{2(X_b')^2}{X_b}
  - \frac{X_b'}{r} \right)  \bigg]g + 16\pi rX_b^2A_b^2\left(1
  - \frac{8A_b^2}{\sigma_0^2} + \frac{12A_b^4}{\sigma_0^4}
  - 2\frac{A_bA_b'}{rX_b^2}  \right)f'.
\end{align}
\end{widetext}
Before we can solve this system numerically, we must determine its
behavior in the limit $r \rightarrow 0$. Bearing in mind the even
parity of scalar functions across $r=0$ and using our freedom to
rescale the perturbations, we can expand them without loss of
generality as $f = 1 + \frac a 2 r^2 + \mathcal{O}(r^4)$ and $g =
\beta + \frac \gamma 2 r^2 + \mathcal{O}(r^4)$.  Inserting into
Eqs.~\eqref{eq:pulsation1} and \eqref{eq:pulsation2} gives us
\begin{align}
  a =& \frac{1}{3}\left(2 + \frac{2\omega^2 - \chi^2}{\alpha_b(0)^2}
  - \frac{32A_b(0)^2}{\sigma_0^2} + \frac{72A_b(0)^4}{\sigma_0^4} \right)
  \nonumber \\
  &- \frac{\gamma}{8\pi A_b(0)^2},
\end{align}
as well as $\beta = 0$ at leading order. There is a degeneracy,
however, at the following order so that $\gamma$ remains undetermined.
In consequence, we have {\it two} undetermined parameters, $\chi$
and $\gamma$, which need to be finetuned to obtain regular spacetimes.

To do this, we need two appropriate boundary conditions at $r
\rightarrow \infty$. Asymptotic flatness requires that $\delta\lambda$
vanishes at infinity. The second condition is obtained by enforcing
that the perturbation conserves the Noether charge
\begin{equation*}
   \mathcal{N} =  \int \sqrt{-\det g}\,J^0\du ^3x\,,
   ~~~J^{\mu} = \frac{\iu}{2}g^{\mu\nu}(\bar{\varphi}\partial_{\nu}
   \varphi-\varphi\partial_{\nu}\bar{\varphi})\,,
 \end{equation*}
compliant with the scalar field's $U(1)$ symmetry.  The variation
in the Noether charge is given by
\begin{equation}
  \delta \mathcal{N} = 4\pi\int_0^\infty \frac{A_b^2r^2}{\alpha_b}dr\left[
  \omega X_b(\delta\lambda - \delta\nu) + 2\delta\dot\psi_2 + 4\omega
  \delta\psi_1 \right]\,.
  \nonumber
\end{equation}
Combining with Eq.~\eqref{eq:lambda_r} allows us to eliminate $\delta
\nu$ whence, using the background equations
\eqref{eq:bg_X}-\eqref{eq:bg_A}, we can write the integrand as a
total derivative and find
\begin{equation}
  \delta \mathcal{N} = \frac{1}{\omega X_b}\left(8\pi r^2\alpha_bA_bA_b'f
  - r\alpha_bg \right)\Big\rvert^{\infty}_{r=0}.
\end{equation}
Despite appearances, the first term is nonnegligible in this
expression due to the exponential behavior of $f$ which vanishes
only for precise choices of the shooting parameters $\chi$ and
$\gamma$.  Thus, for regular perturbations which remain finite at
$r = 0$, the condition $\delta N = 0$ supplies a second boundary
condition at infinity.  Finally, it is worth noting that because
$\delta \lambda$ is related to the perturbation of the total mass,
\begin{equation}
  \delta\lambda = \frac{2X_b^2}{r}\delta M,
\end{equation}
at large radius, our two conditions amount to enforcing that both
the mass and particle number of the BS are unchanged by the
perturbation.  In practice, we solve  Eqs.~\eqref{eq:pulsation1}
and \eqref{eq:pulsation2} using a standard fourth-order Runge-Kutta
method, shooting for the undetermined values $\chi$ and $\gamma$
for each model. We determine these two parameters using a bisection
method similar to that commonly used to determine the frequency
$\omega$ for equilibrium BS models.  Interestingly, we encounter
the same challenging accuracy requirements in our determination of
$\chi$ and $\gamma$ for thin-shell BSs as observed when calculating
the BS frequency $\omega$; therefore we also compute the radial
oscillation spectrum using quadruple precision.

With this method, we can compute real and imaginary $\chi_0$ as
well as overtone frequencies. In this work, we wish to identify
stable systems and therefore restrict our attention to identifying
regimes of positive versus negative $\chi_0^2$.  The point in the
BS parameter space at which models transition from stable to unstable
is marked by the existence of a zero-frequency perturbation, $\chi_0^2
= 0$. Such zero-frequency perturbations can only exist where two
neighboring solutions with the same mass (or, equivalently,  Noether
charge) exist. The condition $\du M / \du A_0 = 0$ is therefore a
necessary but insufficient criterion for the transition of a radial
oscillation mode from stability to instability. For mini BSs, there
exists only one stable branch consisting of models with central
amplitude smaller than that corresponding to the maximum-mass model
\cite{SMGleiser:1988ih}. However, for solitonic models with sufficiently
small $\sigma_0$, a second stable branch emerges; see e.g. Fig.~4
in Ref.~\cite{SMGe:2024itl} for representative examples.

\begin{figure}[t]
  \centering
  \includegraphics[width=\linewidth,clip=true]{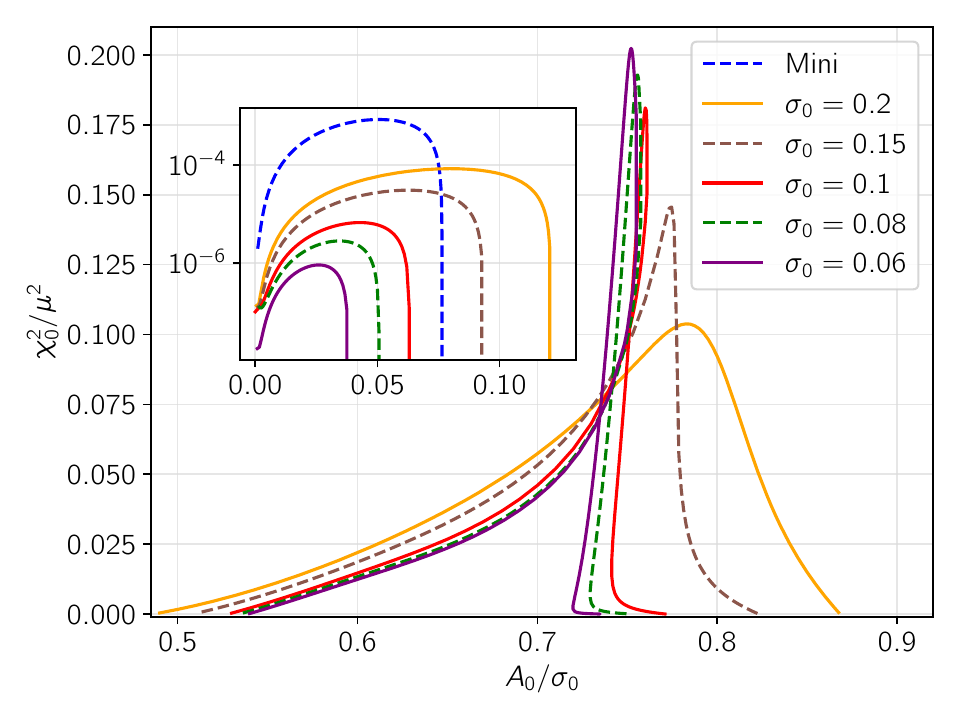}
  \caption{Fundamental radial oscillation frequencies $\chi_0^2$
  against central amplitude $A_0$ for mini and solitonic BSs with
  a variety of $\sigma_0$-values. The main figure shows the second
  stable branch, while the inset shows the first. Central amplitudes
  are rescaled by the appropriate $\sigma_0$-value, apart from mini
  BSs, which undergo no rescaling.
  }
\label{fig:radial_freqs}
\end{figure}

We illustrate the existence of two stable branches for solitonic
BSs in Fig.~\ref{fig:radial_freqs}. There, for each family of BSs,
a positive $\chi_0^2>0$ is obtained across two disjoint regions in
the parameter space, including the ultracompact thin-shell models
in the regime where $M$ and $\omega$ are multi-valued functions of
$A_0$.  Indeed,  the curve $\chi^2_0(A_0)$ also becomes multivalued
for these $\sigma_0$-values, and  self-intersects. In the case of
mini BSs, our results are compatible with those computed in
Ref.~\cite{SMKain:2021rmk}.

\section{Adiabatic Effective Potential}
\label{sup:Adiabatic}

To assess whether our BSs remains ultracompact throughout their
evolution, we need to check whether the light-ring structure persists.
As the BS is evolved in 3+1 dimensions, the spacetime generally
loses its spherical symmetry as a result of numerical errors and
gauge effects that excite asymmetric modes. In terms of the 3+1
metric
\begin{equation} \label{eq:3+1metric}
  \du s^2 = -\alpha^2 \du t^2 + \gamma_{ij}\left(\du x^i + \beta^i \du t
  \right)\left(\du x^j + \beta^j \du t \right)\,,
\end{equation}
where $\alpha, \beta^i$ and $\gamma_{ij}$ are the standard lapse,
shift and spatial metric, this means that the shift will become
non-zero, and the spatial metric will deviate from spherical symmetry.
In principle, the effective potential in the form
\begin{equation}\label{eq: effective potential}
  H(\tilde{r}, \tilde{\theta}) =
  \sqrt{\frac{ - g_{\tilde{t}\tilde{t}}}{g_{\tilde{\phi}\tilde{\phi}}}}\,
\end{equation}
is only defined for stationary spacetimes that exhibit spherical
symmetry; here a tilde distinguishes coordinates of such a spacetime
from those used in our approximately symmetric numerical evolutions.

Given that our evolutions are expected to approximately respect the
symmetries of the initial data, the presence of (stable) bound
geodesics can still be deduced from an Adiabatic Effective Potential
(AEP).  To derive the AEP, we adapt the approach of
Ref.~\cite{SMCunha:2022gde} from (approximate) axisymmetry and apply
it to our spherically symmetric setting.  Under the assumptions
outlined in that work, the metric component $g_{\tilde{t}\tilde{t}}$
in the effective potential \eqref{eq: effective potential} can be
replaced by the evolution variables of the 3+1 line element
\eqref{eq:3+1metric} according to
\begin{equation}\label{eq: AEP t}
  g_{\tilde{t}\tilde{t}} ~~\rightarrow~~ -\langle\alpha^2 \rangle+
  \langle\gamma_{ij}\beta^i \beta^j\rangle.
\end{equation}
Here the brackets denote averaging over a sphere of fixed coordinate
radius $R = \sqrt{x^2+y^2+z^2}$ and $x,y,z$ are the Cartesian
coordinates of the evolution grid.  We convert $R$ into the areal
radius $r$ by calculating the proper area $\mathcal{A}$ of the
sphere\footnote{Note that it is not guaranteed that the coordinate
sphere actually corresponds to a sphere in terms of the areal radius.
Again, if the evolution remains sufficiently close to spherical
symmetry, this is a good approximation.}
\begin{equation}\label{eq: AEP r}
  r^2 = \frac{\mathcal{A}}{4\pi} = \frac{1}{4\pi}\oint_{S^2} \sqrt{\det q}
  \: \du\theta\: \du \phi\,,
\end{equation}
where $q$ is the metric induced on the sphere.  Mapping between
Cartesian coordinates $(x,y,z)$ and angular coordinates $(R,\theta,
\phi)$, the induced metric becomes
\begin{align*}
  q_{\theta\theta} =\, & z^2(\gamma_{xx}\cos^2 \phi + \gamma_{yy}\sin^2 \phi)
  + \gamma_{zz} R^2 \sin^2\theta \\
  & +\gamma_{xy}z^2\sin(2\phi) -
  2zR\sin\theta(\gamma_{xz}\cos\phi + \gamma_{yz}\sin\phi)\,,\\
  q_{\phi\phi} =\, & \gamma_{xx} y^2 - 2\gamma_{xy}xy + \gamma_{yy}x^2\,,\\
  q_{\theta\phi} = \, & \gamma_{yy}xz \sin\phi -\gamma_{xx}yz \cos\phi+
  \gamma_{xy} z(x \cos\phi - y \sin\phi) \\
  & + R \sin\theta(\gamma_{xz} y- \gamma_{yz} x)\,.
\end{align*}
In the equatorial plane $\tilde{\theta} = \frac\pi2$ of the stationary
and spherically symmetric spacetime, we have $g_{\tilde{\phi}\tilde{\phi}}
= \tilde{r}^2$.  Replacing $\tilde{r} \rightarrow r$ and using
Eqs.~\eqref{eq: effective potential}-\eqref{eq: AEP r} we now define
the AEP as
\begin{equation}\label{eq: AEP}
  H_{\text{eff}} (r) = \sqrt{\frac{\langle\alpha^2 \rangle
  - \langle\gamma_{ij}\beta^i \beta^j\rangle}{r}}\,.
\end{equation}
In the idealized case where our simulation is perfectly spherically
symmetric and stationary, Eq.~\eqref{eq: AEP} reduces to Eq.~\eqref{eq:
effective potential}. As long as this AEP has extrema we have
evidence that the BS is still ultracompact, featuring a pair of
(perturbed) light rings. In practice, we extract the AEP at a number
of coordinate radii with resolution $\Delta R \approx 0.1$ in the
range where we expect the light rings to be.

\begin{figure}
  \centering
  \includegraphics[width=\linewidth]{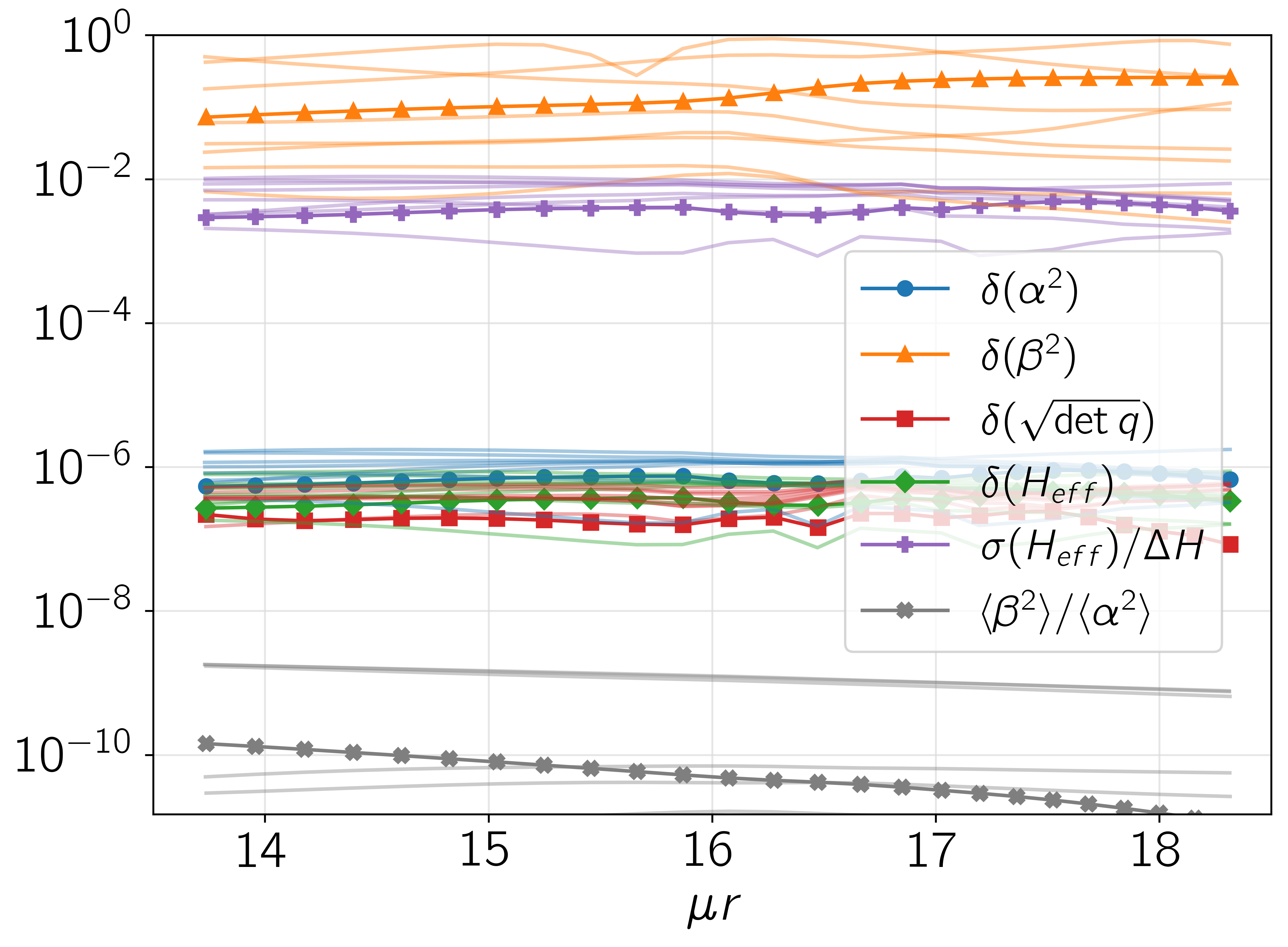}
  \caption{Diagnostics to assess the deviations from spherical
  symmetry and stationarity for \texttt{S08A06}. Brackets denote
  averaging over an extraction sphere, whereas $\sigma$ denotes the
  standard deviation from this average value. The relative standard
  deviation is $\delta(X) \equiv \sigma(X) / \langle X\rangle$, and
  $\Delta H$ denotes the potential difference between the two extrema
  of the AEP. The lines with markers correspond to time $\mu t =
  75$, after early gauge dynamics have settled down. The other lines
  correspond to extractions at later times ($\mu t \in \{250,500,
  \dots, 2000\}$) to illustrate that the conclusions also hold at
  later times.}
  \label{fig: sphere variances s08}
\end{figure}

In order to quantify the deviation from spherical symmetry, we
compute the variation of the relevant quantities over the extraction
spheres. For this purpose, we define the relative standard deviation
of a variable $X$ as
\begin{equation} \label{eq:standard_deviation}
  \delta (X) \equiv \frac{\sigma(X)}{\langle X \rangle} \equiv
  \frac{\sqrt{\langle\left(X - \langle X\rangle\right)^2\rangle}}{\langle X \rangle}\,,
\end{equation}
Figure \ref{fig: sphere variances s08} shows the relative standard
deviation for the individual terms of Eq.~\eqref{eq: AEP}, as well
as that of the AEP itself, as obtained for 192 points on each sphere.

The largest relative deviation from spherical symmetry is seen for
$\beta^2 \equiv \gamma_{ij}\beta^i\beta^j$, where $\delta(\beta^2)
\gtrsim 10^{-2}$, reaching order unity at some times; this is merely
a consequence of our definition \eqref{eq:standard_deviation} and
the near-zero average $\langle \beta^2 \rangle$.  This interpretation
is supported by the very small ratio $\langle \beta^2 \rangle /
\langle \alpha^2 \rangle \lesssim 10^{-9}$. The relative standard
deviation of the remaining variables is $\lesssim 10^{-6}$, suggesting
that the spacetime is very close to spherical symmetry and approximately
stationary.  We further investigate whether these small variations
have any influence on the light rings' presence and structure.  For
this purpose we also display $\sigma(H_{\text{eff}}) / \Delta H$
as the purple curve in Fig.~\ref{fig: sphere variances s08}, where
$\Delta H$ is the potential difference between the extracted extrema
of the AEP. Its value varies around $\sim 10^{-3} - 10^{-2}$,
implying that the deviations from our symmetry assumptions explain
at most 1\% of the observed potential difference between the extrema.

\begin{figure}[t]
  \centering
  \includegraphics[width=\linewidth]{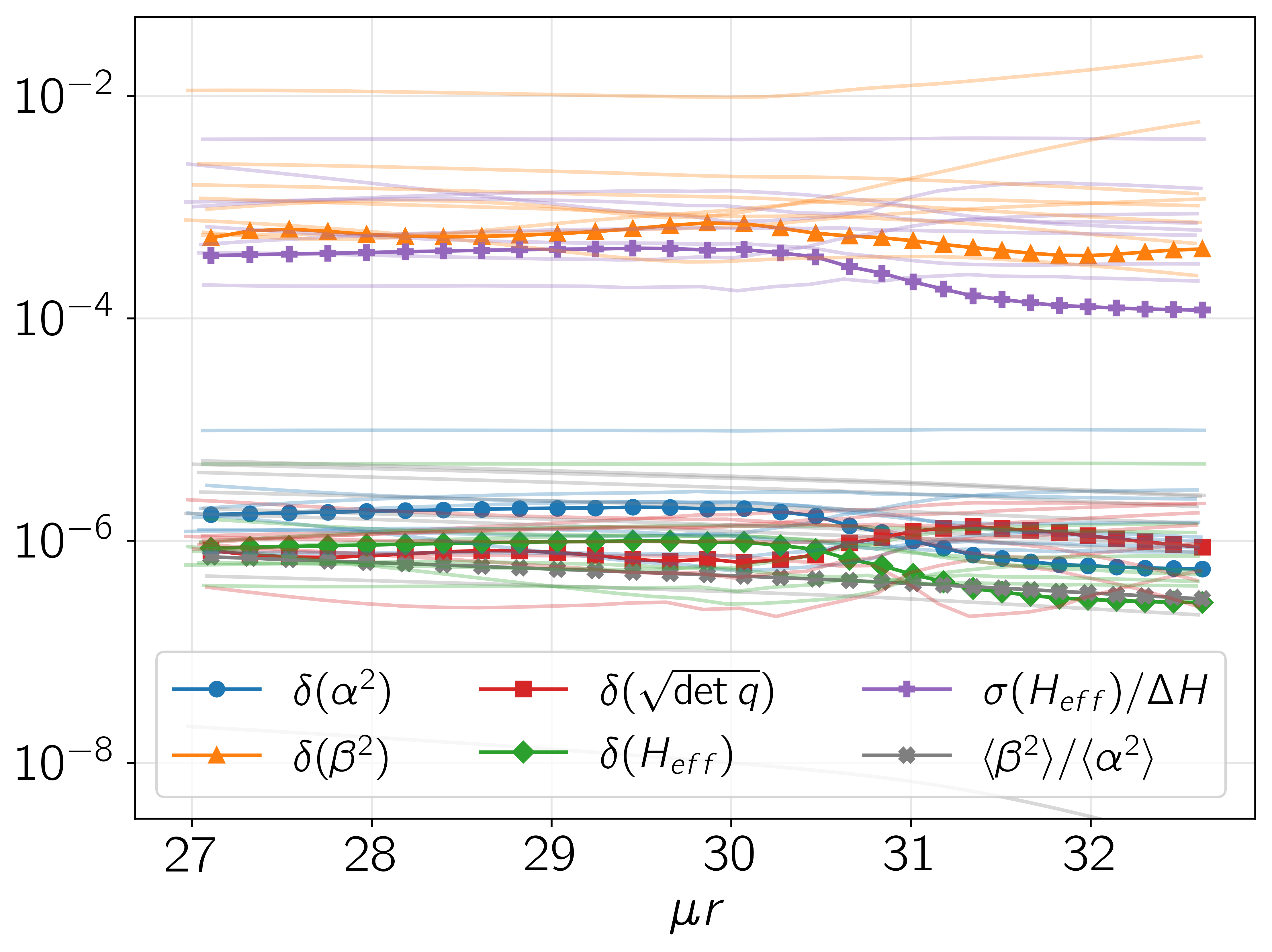}
  \caption{Same as Fig.~\ref{fig: sphere variances s08}, but for BS model
  \texttt{S06A044}.}
  \label{fig: sphere variances s06}
\end{figure}

The conclusions for \texttt{S06A044} are similar and displayed in
Fig.~\ref{fig: sphere variances s06}. The major difference is that
$\langle\beta^2\rangle / \langle\alpha^2 \rangle \sim 10^{-6}$ is
considerably larger now. This is in line with the more pronounced
oscillations also observed in other diagnostics for this numerically
more challenging configuration; cf.~Figs.~2 and~4 in the main text.
The ratio $\sigma(H_{\text{eff}}) / \Delta H \lesssim 10^{-3}$,
however, is smaller by an order of magnitude, a side effect of the
more pronounced extrema in the potential for this more extreme
model.

\section{Ray tracing}

The ray-traced images presented in this work (cf.Fig.~6 of the main
text) are obtained using Flexible Object-Oriented Ray Tracer (FOORT),
which is publicly available on GitHub \cite{SMnewMayerson:2024fgh,
SMMayerson:2025foo}. The geodesic equations are integrated using the
Runge-Kutta 4 or the Verlet scheme to produce images of a variety
of spacetimes. FOORT is highly flexible and easily accommodates new
analytical or numerical spacetimes. It employs adaptive mesh
refinement (AMR) by refining the initially uniform grid in regions
where selected image features change significantly.

The images in the main text are produced for the \texttt{S06A044}
BS model using the Verlet scheme, for an observer located at areal
radius $\mu r = 1000$ in units where the mass of the boson star is
$\mu M \approx 10.4$. The image region has dimensions $150 \times
150$ in these units, and the highest resolution obtained with the
AMR is $\sim 0.047$. The insets focus on a region of $7.5\times7.5$,
with highest resolution $\sim 0.0047$.  The emission is modeled
through an optically thin equatorial disk with emission profile
given by~\cite{SMGralla:2020sha}
\begin{align}
  I_0(r) = \,& \left[(r-\rho)^2+\Delta^2\right]^{-1/2} \nonumber \\
  & \times \exp\left\{-\frac12 \left[\kappa + \operatorname{arcsinh}
  \left(\frac{r-\rho}{\Delta}\right)\right]^2\right\}\,.
  \label{eq: GLM emission}
\end{align}
The parameters used are $(\rho, \Delta, \kappa) = (60, 2.5, -2)$,
with the first value corresponding to $\sim 6M$, the radial coordinate
of the ISCO in Schwarzschild spacetime. Although the spacetime is
non-rotating, the fluid velocity can be modeled with a mix of radial
and (sub)Keplerian orbits, as described in Refs.~\cite{SMPu:2016flo,
SMCardenasAvendano:2023aar}. For the images in the main text, parameters
$\xi = \beta_r = \beta_\phi = 1.0$ (see Eqs.~(B36),~(B50),~(B51)
in Ref.~\cite{SMCardenasAvendano:2023aar}) have been used, corresponding
to pure Keplerian orbits which give rise to the observed asymmetry
in the images.

%


\begin{thebibliography}{49}%
\makeatletter
\providecommand \@ifxundefined [1]{%
 \@ifx{#1\undefined}
}%
\providecommand \@ifnum [1]{%
 \ifnum #1\expandafter \@firstoftwo
 \else \expandafter \@secondoftwo
 \fi
}%
\providecommand \@ifx [1]{%
 \ifx #1\expandafter \@firstoftwo
 \else \expandafter \@secondoftwo
 \fi
}%
\providecommand \natexlab [1]{#1}%
\providecommand \enquote  [1]{``#1''}%
\providecommand \bibnamefont  [1]{#1}%
\providecommand \bibfnamefont [1]{#1}%
\providecommand \citenamefont [1]{#1}%
\providecommand \href@noop [0]{\@secondoftwo}%
\providecommand \href [0]{\begingroup \@sanitize@url \@href}%
\providecommand \@href[1]{\@@startlink{#1}\@@href}%
\providecommand \@@href[1]{\endgroup#1\@@endlink}%
\providecommand \@sanitize@url [0]{\catcode `\\12\catcode `\$12\catcode
  `\&12\catcode `\#12\catcode `\^12\catcode `\_12\catcode `\%12\relax}%
\providecommand \@@startlink[1]{}%
\providecommand \@@endlink[0]{}%
\providecommand \url  [0]{\begingroup\@sanitize@url \@url }%
\providecommand \@url [1]{\endgroup\@href {#1}{\urlprefix }}%
\providecommand \urlprefix  [0]{URL }%
\providecommand \Eprint [0]{\href }%
\providecommand \doibase [0]{http://dx.doi.org/}%
\providecommand \selectlanguage [0]{\@gobble}%
\providecommand \bibinfo  [0]{\@secondoftwo}%
\providecommand \bibfield  [0]{\@secondoftwo}%
\providecommand \translation [1]{[#1]}%
\providecommand \BibitemOpen [0]{}%
\providecommand \bibitemStop [0]{}%
\providecommand \bibitemNoStop [0]{.\EOS\space}%
\providecommand \EOS [0]{\spacefactor3000\relax}%
\providecommand \BibitemShut  [1]{\csname bibitem#1\endcsname}%
\let\auto@bib@innerbib\@empty
\bibitem [{\citenamefont {Bowyer}\ \emph {et~al.}(1965)\citenamefont {Bowyer},
  \citenamefont {Byram}, \citenamefont {Chubb},\ and\ \citenamefont
  {Friedmann}}]{Bowyer:1965}%
  \BibitemOpen
  \bibfield  {author} {\bibinfo {author} {\bibfnamefont {S.}~\bibnamefont
  {Bowyer}}, \bibinfo {author} {\bibfnamefont {E.~T.}\ \bibnamefont {Byram}},
  \bibinfo {author} {\bibfnamefont {T.~A.}\ \bibnamefont {Chubb}}, \ and\
  \bibinfo {author} {\bibfnamefont {H.}~\bibnamefont {Friedmann}},\ }\href
  {\doibase 10.1126/science.147.3656.394} {\bibfield  {journal} {\bibinfo
  {journal} {Science}\ }\textbf {\bibinfo {volume} {147}},\ \bibinfo {pages}
  {394} (\bibinfo {year} {1965})}\BibitemShut {NoStop}%
\bibitem [{\citenamefont {Hazard}\ \emph {et~al.}(1963)\citenamefont {Hazard},
  \citenamefont {Mackey},\ and\ \citenamefont {Shimmins}}]{Hazard:1963}%
  \BibitemOpen
  \bibfield  {author} {\bibinfo {author} {\bibfnamefont {C.}~\bibnamefont
  {Hazard}}, \bibinfo {author} {\bibfnamefont {M.~B.}\ \bibnamefont {Mackey}},
  \ and\ \bibinfo {author} {\bibfnamefont {A.~J.}\ \bibnamefont {Shimmins}},\
  }\href {\doibase doi.org/10.1038/1971037a0} {\bibfield  {journal} {\bibinfo
  {journal} {Nature}\ }\textbf {\bibinfo {volume} {197}},\ \bibinfo {pages}
  {1037} (\bibinfo {year} {1963})}\BibitemShut {NoStop}%
\bibitem [{\citenamefont {Schmidt}(1963)}]{Schmidt:1963}%
  \BibitemOpen
  \bibfield  {author} {\bibinfo {author} {\bibfnamefont {M.}~\bibnamefont
  {Schmidt}},\ }\href {\doibase 10.1038/1971040a0} {\bibfield  {journal}
  {\bibinfo  {journal} {Nature}\ }\textbf {\bibinfo {volume} {197}},\ \bibinfo
  {pages} {1040} (\bibinfo {year} {1963})}\BibitemShut {NoStop}%
\bibitem [{\citenamefont {Abbott}\ \emph {et~al.}(2016)\citenamefont {Abbott}
  \emph {et~al.}}]{Abbott:2016blz}%
  \BibitemOpen
  \bibfield  {author} {\bibinfo {author} {\bibfnamefont {B.~P.}\ \bibnamefont
  {Abbott}} \emph {et~al.},\ }\href {\doibase 10.1103/PhysRevLett.116.061102}
  {\bibfield  {journal} {\bibinfo  {journal} {Phys. Rev. Lett.}\ }\textbf
  {\bibinfo {volume} {116}},\ \bibinfo {pages} {061102} (\bibinfo {year}
  {2016})},\ \Eprint {http://arxiv.org/abs/1602.03837} {arXiv:1602.03837
  [gr-qc]} \BibitemShut {NoStop}%
\bibitem [{\citenamefont {Abbott}\ \emph {et~al.}(2023)\citenamefont {Abbott}
  \emph {et~al.}}]{KAGRA:2021vkt}%
  \BibitemOpen
  \bibfield  {author} {\bibinfo {author} {\bibfnamefont {R.}~\bibnamefont
  {Abbott}} \emph {et~al.} (\bibinfo {collaboration} {KAGRA, VIRGO, LIGO
  Scientific}),\ }\href {\doibase 10.1103/PhysRevX.13.041039} {\bibfield
  {journal} {\bibinfo  {journal} {Phys. Rev. X}\ }\textbf {\bibinfo {volume}
  {13}},\ \bibinfo {pages} {041039} (\bibinfo {year} {2023})},\ \Eprint
  {http://arxiv.org/abs/2111.03606} {arXiv:2111.03606 [gr-qc]} \BibitemShut
  {NoStop}%
\bibitem [{\citenamefont {Abbott}\ \emph {et~al.}(2021)\citenamefont {Abbott}
  \emph {et~al.}}]{LIGOScientific:2021sio}%
  \BibitemOpen
  \bibfield  {author} {\bibinfo {author} {\bibfnamefont {R.}~\bibnamefont
  {Abbott}} \emph {et~al.} (\bibinfo {collaboration} {LIGO Scientific, VIRGO,
  KAGRA}),\ }\href@noop {} {\  (\bibinfo {year} {2021})},\ \Eprint
  {http://arxiv.org/abs/2112.06861} {arXiv:2112.06861 [gr-qc]} \BibitemShut
  {NoStop}%
\bibitem [{\citenamefont {Akiyama}\ \emph {et~al.}(2019)\citenamefont {Akiyama}
  \emph {et~al.}}]{EventHorizonTelescope:2019dse}%
  \BibitemOpen
  \bibfield  {author} {\bibinfo {author} {\bibfnamefont {K.}~\bibnamefont
  {Akiyama}} \emph {et~al.} (\bibinfo {collaboration} {Event Horizon
  Telescope}),\ }\href {\doibase 10.3847/2041-8213/ab0ec7} {\bibfield
  {journal} {\bibinfo  {journal} {Astrophys. J. Lett.}\ }\textbf {\bibinfo
  {volume} {875}},\ \bibinfo {pages} {L1} (\bibinfo {year} {2019})},\ \Eprint
  {http://arxiv.org/abs/1906.11238} {arXiv:1906.11238 [astro-ph.GA]}
  \BibitemShut {NoStop}%
\bibitem [{\citenamefont {Akiyama}\ \emph {et~al.}(2022)\citenamefont {Akiyama}
  \emph {et~al.}}]{EventHorizonTelescope:2022wkp}%
  \BibitemOpen
  \bibfield  {author} {\bibinfo {author} {\bibfnamefont {K.}~\bibnamefont
  {Akiyama}} \emph {et~al.} (\bibinfo {collaboration} {Event Horizon
  Telescope}),\ }\href {\doibase 10.3847/2041-8213/ac6674} {\bibfield
  {journal} {\bibinfo  {journal} {Astrophys. J. Lett.}\ }\textbf {\bibinfo
  {volume} {930}},\ \bibinfo {pages} {L12} (\bibinfo {year} {2022})},\ \Eprint
  {http://arxiv.org/abs/2311.08680} {arXiv:2311.08680 [astro-ph.HE]}
  \BibitemShut {NoStop}%
\bibitem [{\citenamefont {Synge}(1966)}]{Synge:1966okc}%
  \BibitemOpen
  \bibfield  {author} {\bibinfo {author} {\bibfnamefont {J.~L.}\ \bibnamefont
  {Synge}},\ }\href {\doibase 10.1093/mnras/131.3.463} {\bibfield  {journal}
  {\bibinfo  {journal} {Mon. Not. Roy. Astron. Soc.}\ }\textbf {\bibinfo
  {volume} {131}},\ \bibinfo {pages} {463} (\bibinfo {year}
  {1966})}\BibitemShut {NoStop}%
\bibitem [{\citenamefont {Cardoso}\ \emph {et~al.}(2009)\citenamefont
  {Cardoso}, \citenamefont {Miranda}, \citenamefont {Berti}, \citenamefont
  {Witek},\ and\ \citenamefont {Zanchin}}]{Cardoso:2008bp}%
  \BibitemOpen
  \bibfield  {author} {\bibinfo {author} {\bibfnamefont {V.}~\bibnamefont
  {Cardoso}}, \bibinfo {author} {\bibfnamefont {A.~S.}\ \bibnamefont
  {Miranda}}, \bibinfo {author} {\bibfnamefont {E.}~\bibnamefont {Berti}},
  \bibinfo {author} {\bibfnamefont {H.}~\bibnamefont {Witek}}, \ and\ \bibinfo
  {author} {\bibfnamefont {V.~T.}\ \bibnamefont {Zanchin}},\ }\href {\doibase
  10.1103/PhysRevD.79.064016} {\bibfield  {journal} {\bibinfo  {journal} {Phys.
  Rev. D}\ }\textbf {\bibinfo {volume} {79}},\ \bibinfo {pages} {064016}
  (\bibinfo {year} {2009})},\ \bibinfo {note} {arXiv:0812.1806
  [hep-th]}\BibitemShut {NoStop}%
\bibitem [{\citenamefont {Jusufi}(2020)}]{Jusufi:2019ltj}%
  \BibitemOpen
  \bibfield  {author} {\bibinfo {author} {\bibfnamefont {K.}~\bibnamefont
  {Jusufi}},\ }\href {\doibase 10.1103/PhysRevD.101.084055} {\bibfield
  {journal} {\bibinfo  {journal} {Phys. Rev. D}\ }\textbf {\bibinfo {volume}
  {101}},\ \bibinfo {pages} {084055} (\bibinfo {year} {2020})},\ \Eprint
  {http://arxiv.org/abs/1912.13320} {arXiv:1912.13320 [gr-qc]} \BibitemShut
  {NoStop}%
\bibitem [{\citenamefont {Koga}\ \emph {et~al.}(2022)\citenamefont {Koga},
  \citenamefont {Asaka}, \citenamefont {Kimura},\ and\ \citenamefont
  {Okabayashi}}]{Koga:2022dsu}%
  \BibitemOpen
  \bibfield  {author} {\bibinfo {author} {\bibfnamefont {Y.}~\bibnamefont
  {Koga}}, \bibinfo {author} {\bibfnamefont {N.}~\bibnamefont {Asaka}},
  \bibinfo {author} {\bibfnamefont {M.}~\bibnamefont {Kimura}}, \ and\ \bibinfo
  {author} {\bibfnamefont {K.}~\bibnamefont {Okabayashi}},\ }\href {\doibase
  10.1103/PhysRevD.105.104040} {\bibfield  {journal} {\bibinfo  {journal}
  {Phys. Rev. D}\ }\textbf {\bibinfo {volume} {105}},\ \bibinfo {pages}
  {104040} (\bibinfo {year} {2022})},\ \Eprint
  {http://arxiv.org/abs/2202.00201} {arXiv:2202.00201 [gr-qc]} \BibitemShut
  {NoStop}%
\bibitem [{\citenamefont {V\"olkel}\ \emph {et~al.}(2022)\citenamefont
  {V\"olkel}, \citenamefont {Franchini}, \citenamefont {Barausse},\ and\
  \citenamefont {Berti}}]{Volkel:2022khh}%
  \BibitemOpen
  \bibfield  {author} {\bibinfo {author} {\bibfnamefont {S.~H.}\ \bibnamefont
  {V\"olkel}}, \bibinfo {author} {\bibfnamefont {N.}~\bibnamefont {Franchini}},
  \bibinfo {author} {\bibfnamefont {E.}~\bibnamefont {Barausse}}, \ and\
  \bibinfo {author} {\bibfnamefont {E.}~\bibnamefont {Berti}},\ }\href
  {\doibase 10.1103/PhysRevD.106.124036} {\bibfield  {journal} {\bibinfo
  {journal} {Phys. Rev. D}\ }\textbf {\bibinfo {volume} {106}},\ \bibinfo
  {pages} {124036} (\bibinfo {year} {2022})},\ \Eprint
  {http://arxiv.org/abs/2209.10564} {arXiv:2209.10564 [gr-qc]} \BibitemShut
  {NoStop}%
\bibitem [{\citenamefont {Pedrotti}\ and\ \citenamefont
  {Calz{\`a}}(2025)}]{Pedrotti:2025idg}%
  \BibitemOpen
  \bibfield  {author} {\bibinfo {author} {\bibfnamefont {D.}~\bibnamefont
  {Pedrotti}}\ and\ \bibinfo {author} {\bibfnamefont {M.}~\bibnamefont
  {Calz{\`a}}},\ }\href {\doibase 10.1103/1q35-mjjz} {\bibfield  {journal}
  {\bibinfo  {journal} {Phys. Rev. D}\ }\textbf {\bibinfo {volume} {111}},\
  \bibinfo {pages} {124056} (\bibinfo {year} {2025})},\ \Eprint
  {http://arxiv.org/abs/2504.01909} {arXiv:2504.01909 [gr-qc]} \BibitemShut
  {NoStop}%
\bibitem [{\citenamefont {Hawking}\ and\ \citenamefont
  {Ellis}(1973)}]{Hawking:1973uf}%
  \BibitemOpen
  \bibfield  {author} {\bibinfo {author} {\bibfnamefont {S.~W.}\ \bibnamefont
  {Hawking}}\ and\ \bibinfo {author} {\bibfnamefont {G.~F.~R.}\ \bibnamefont
  {Ellis}},\ }\href@noop {} {\emph {\bibinfo {title} {{The Large Scale
  Structure of Space-Time}}}}\ (\bibinfo  {publisher} {Cambridge University
  Press},\ \bibinfo {year} {1973})\BibitemShut {NoStop}%
\bibitem [{\citenamefont {Mathur}(2009)}]{Mathur:2009ip}%
  \BibitemOpen
  \bibfield  {author} {\bibinfo {author} {\bibfnamefont {S.~D.}\ \bibnamefont
  {Mathur}},\ }\href {\doibase 10.1088/0264-9381/26/22/224001} {\bibfield
  {journal} {\bibinfo  {journal} {Classical and Quantum Gravity}\ }\textbf
  {\bibinfo {volume} {26}},\ \bibinfo {pages} {224001} (\bibinfo {year}
  {2009})}\BibitemShut {NoStop}%
\bibitem [{\citenamefont {Evstafyeva}\ \emph {et~al.}(2024)\citenamefont
  {Evstafyeva}, \citenamefont {Sperhake}, \citenamefont {Romero-Shaw},\ and\
  \citenamefont {Agathos}}]{Evstafyeva:2024qvp}%
  \BibitemOpen
  \bibfield  {author} {\bibinfo {author} {\bibfnamefont {T.}~\bibnamefont
  {Evstafyeva}}, \bibinfo {author} {\bibfnamefont {U.}~\bibnamefont
  {Sperhake}}, \bibinfo {author} {\bibfnamefont {I.}~\bibnamefont
  {Romero-Shaw}}, \ and\ \bibinfo {author} {\bibfnamefont {M.}~\bibnamefont
  {Agathos}},\ }\href {\doibase 10.1103/PhysRevLett.133.131401} {\bibfield
  {journal} {\bibinfo  {journal} {Phys. Rev. Lett.}\ }\textbf {\bibinfo
  {volume} {133}},\ \bibinfo {pages} {131401} (\bibinfo {year} {2024})},\
  \bibinfo {note} {arXiv:2406.02715 [gr-qc]},\ \Eprint
  {http://arxiv.org/abs/2406.02715} {arXiv:2406.02715 [gr-qc]} \BibitemShut
  {NoStop}%
\bibitem [{\citenamefont {Cunha}\ \emph {et~al.}(2017)\citenamefont {Cunha},
  \citenamefont {Berti},\ and\ \citenamefont {Herdeiro}}]{Cunha:2017qtt}%
  \BibitemOpen
  \bibfield  {author} {\bibinfo {author} {\bibfnamefont {P.~V.~P.}\
  \bibnamefont {Cunha}}, \bibinfo {author} {\bibfnamefont {E.}~\bibnamefont
  {Berti}}, \ and\ \bibinfo {author} {\bibfnamefont {C.~A.~R.}\ \bibnamefont
  {Herdeiro}},\ }\href {\doibase 10.1103/PhysRevLett.119.251102} {\bibfield
  {journal} {\bibinfo  {journal} {Phys. Rev. Lett.}\ }\textbf {\bibinfo
  {volume} {119}},\ \bibinfo {pages} {251102} (\bibinfo {year} {2017})},\
  \Eprint {http://arxiv.org/abs/arXiv:1708.04211 [gr-qc]} {arXiv:1708.04211
  [gr-qc]} \BibitemShut {NoStop}%
\bibitem [{\citenamefont {Cunha}\ and\ \citenamefont
  {Herdeiro}(2020)}]{Cunha:2020azh}%
  \BibitemOpen
  \bibfield  {author} {\bibinfo {author} {\bibfnamefont {P.~V.~P.}\
  \bibnamefont {Cunha}}\ and\ \bibinfo {author} {\bibfnamefont {C.~A.~R.}\
  \bibnamefont {Herdeiro}},\ }\href {\doibase 10.1103/PhysRevLett.124.181101}
  {\bibfield  {journal} {\bibinfo  {journal} {Phys. Rev. Lett.}\ }\textbf
  {\bibinfo {volume} {124}},\ \bibinfo {pages} {181101} (\bibinfo {year}
  {2020})},\ \Eprint {http://arxiv.org/abs/2003.06445} {arXiv:2003.06445
  [gr-qc]} \BibitemShut {NoStop}%
\bibitem [{\citenamefont {Keir}(2016)}]{Keir:2014oka}%
  \BibitemOpen
  \bibfield  {author} {\bibinfo {author} {\bibfnamefont {J.}~\bibnamefont
  {Keir}},\ }\href {\doibase 10.1088/0264-9381/33/13/135009} {\bibfield
  {journal} {\bibinfo  {journal} {Class. Quant. Grav.}\ }\textbf {\bibinfo
  {volume} {33}},\ \bibinfo {pages} {135009} (\bibinfo {year} {2016})},\
  \Eprint {http://arxiv.org/abs/1404.7036} {arXiv:1404.7036 [gr-qc]}
  \BibitemShut {NoStop}%
\bibitem [{\citenamefont {Bizo{\'n}}\ and\ \citenamefont
  {Rostworowski}(2011)}]{Bizon:2011gg}%
  \BibitemOpen
  \bibfield  {author} {\bibinfo {author} {\bibfnamefont {P.}~\bibnamefont
  {Bizo{\'n}}}\ and\ \bibinfo {author} {\bibfnamefont {A.}~\bibnamefont
  {Rostworowski}},\ }\href {\doibase 10.1103/PhysRevLett.107.031102} {\bibfield
   {journal} {\bibinfo  {journal} {Phys. Rev. Lett.}\ }\textbf {\bibinfo
  {volume} {107}},\ \bibinfo {pages} {031102} (\bibinfo {year} {2011})},\
  \bibinfo {note} {arXiv:1104.3702 [gr-qc]}\BibitemShut {NoStop}%
\bibitem [{\citenamefont {Holzegel}\ and\ \citenamefont
  {Smulevici}(2013)}]{Holzegel:2011uu}%
  \BibitemOpen
  \bibfield  {author} {\bibinfo {author} {\bibfnamefont {G.}~\bibnamefont
  {Holzegel}}\ and\ \bibinfo {author} {\bibfnamefont {J.}~\bibnamefont
  {Smulevici}},\ }\href {\doibase 10.1002/cpa.21470} {\bibfield  {journal}
  {\bibinfo  {journal} {Commun. Pure Appl. Math.}\ }\textbf {\bibinfo {volume}
  {66}},\ \bibinfo {pages} {1751} (\bibinfo {year} {2013})},\ \Eprint
  {http://arxiv.org/abs/1110.6794} {arXiv:1110.6794 [gr-qc]} \BibitemShut
  {NoStop}%
\bibitem [{\citenamefont {Holzegel}\ and\ \citenamefont
  {Smulevici}(2014)}]{Holzegel:2013kna}%
  \BibitemOpen
  \bibfield  {author} {\bibinfo {author} {\bibfnamefont {G.}~\bibnamefont
  {Holzegel}}\ and\ \bibinfo {author} {\bibfnamefont {J.}~\bibnamefont
  {Smulevici}},\ }\href {\doibase 10.2140/apde.2014.7.1057} {\bibfield
  {journal} {\bibinfo  {journal} {Anal. Part. Diff. Eq.}\ }\textbf {\bibinfo
  {volume} {7}},\ \bibinfo {pages} {1057} (\bibinfo {year} {2014})},\ \Eprint
  {http://arxiv.org/abs/1303.5944} {arXiv:1303.5944 [gr-qc]} \BibitemShut
  {NoStop}%
\bibitem [{\citenamefont {Cardoso}\ \emph {et~al.}(2014)\citenamefont
  {Cardoso}, \citenamefont {Crispino}, \citenamefont {Macedo}, \citenamefont
  {Okawa},\ and\ \citenamefont {Pani}}]{Cardoso:2014sna}%
  \BibitemOpen
  \bibfield  {author} {\bibinfo {author} {\bibfnamefont {V.}~\bibnamefont
  {Cardoso}}, \bibinfo {author} {\bibfnamefont {L.~C.~B.}\ \bibnamefont
  {Crispino}}, \bibinfo {author} {\bibfnamefont {C.~F.~B.}\ \bibnamefont
  {Macedo}}, \bibinfo {author} {\bibfnamefont {H.}~\bibnamefont {Okawa}}, \
  and\ \bibinfo {author} {\bibfnamefont {P.}~\bibnamefont {Pani}},\ }\href
  {\doibase 10.1103/PhysRevD.90.044069} {\bibfield  {journal} {\bibinfo
  {journal} {Phys. Rev. D}\ }\textbf {\bibinfo {volume} {90}},\ \bibinfo
  {pages} {044069} (\bibinfo {year} {2014})},\ \Eprint
  {http://arxiv.org/abs/1406.5510} {arXiv:1406.5510 [gr-qc]} \BibitemShut
  {NoStop}%
\bibitem [{\citenamefont {Cunha}\ \emph {et~al.}(2023)\citenamefont {Cunha},
  \citenamefont {Herdeiro}, \citenamefont {Radu},\ and\ \citenamefont
  {Sanchis-Gual}}]{Cunha:2022gde}%
  \BibitemOpen
  \bibfield  {author} {\bibinfo {author} {\bibfnamefont {P.~V.~P.}\
  \bibnamefont {Cunha}}, \bibinfo {author} {\bibfnamefont {C.}~\bibnamefont
  {Herdeiro}}, \bibinfo {author} {\bibfnamefont {E.}~\bibnamefont {Radu}}, \
  and\ \bibinfo {author} {\bibfnamefont {N.}~\bibnamefont {Sanchis-Gual}},\
  }\href {\doibase 10.1103/PhysRevLett.130.061401} {\bibfield  {journal}
  {\bibinfo  {journal} {Phys. Rev. Lett.}\ }\textbf {\bibinfo {volume} {130}},\
  \bibinfo {pages} {061401} (\bibinfo {year} {2023})},\ \Eprint
  {http://arxiv.org/abs/2207.13713} {arXiv:2207.13713 [gr-qc]} \BibitemShut
  {NoStop}%
\bibitem [{\citenamefont {Cunha}(2025)}]{Cunha:2025oeu}%
  \BibitemOpen
  \bibfield  {author} {\bibinfo {author} {\bibfnamefont {P.~V.~P.}\
  \bibnamefont {Cunha}},\ }\href@noop {} {\  (\bibinfo {year} {2025})},\
  \Eprint {http://arxiv.org/abs/2503.00117} {arXiv:2503.00117 [gr-qc]}
  \BibitemShut {NoStop}%
\bibitem [{\citenamefont {Siemonsen}(2024)}]{Siemonsen:2024snb}%
  \BibitemOpen
  \bibfield  {author} {\bibinfo {author} {\bibfnamefont {N.}~\bibnamefont
  {Siemonsen}},\ }\href {\doibase 10.1103/PhysRevLett.133.031401} {\bibfield
  {journal} {\bibinfo  {journal} {Phys. Rev. Lett.}\ }\textbf {\bibinfo
  {volume} {133}},\ \bibinfo {pages} {031401} (\bibinfo {year} {2024})},\
  \Eprint {http://arxiv.org/abs/2404.14536} {arXiv:2404.14536 [gr-qc]}
  \BibitemShut {NoStop}%
\bibitem [{\citenamefont {Guo}\ \emph {et~al.}(2024)\citenamefont {Guo},
  \citenamefont {Wang},\ and\ \citenamefont {Zhang}}]{Guo:2024cts}%
  \BibitemOpen
  \bibfield  {author} {\bibinfo {author} {\bibfnamefont {G.}~\bibnamefont
  {Guo}}, \bibinfo {author} {\bibfnamefont {P.}~\bibnamefont {Wang}}, \ and\
  \bibinfo {author} {\bibfnamefont {Y.}~\bibnamefont {Zhang}},\ }\href@noop {}
  {\  (\bibinfo {year} {2024})},\ \Eprint {http://arxiv.org/abs/2403.02089}
  {arXiv:2403.02089 [gr-qc]} \BibitemShut {NoStop}%
\bibitem [{\citenamefont {Benomio}\ \emph {et~al.}(2024)\citenamefont
  {Benomio}, \citenamefont {C\'ardenas-Avenda\~no}, \citenamefont {Pretorius},\
  and\ \citenamefont {Sullivan}}]{Benomio:2024lev}%
  \BibitemOpen
  \bibfield  {author} {\bibinfo {author} {\bibfnamefont {G.}~\bibnamefont
  {Benomio}}, \bibinfo {author} {\bibfnamefont {A.}~\bibnamefont
  {C\'ardenas-Avenda\~no}}, \bibinfo {author} {\bibfnamefont {F.}~\bibnamefont
  {Pretorius}}, \ and\ \bibinfo {author} {\bibfnamefont {A.}~\bibnamefont
  {Sullivan}},\ }\href@noop {} {\  (\bibinfo {year} {2024})},\ \Eprint
  {http://arxiv.org/abs/2411.17445} {arXiv:2411.17445 [gr-qc]} \BibitemShut
  {NoStop}%
\bibitem [{\citenamefont {Redondo-Yuste}\ and\ \citenamefont
  {C\'ardenas-Avenda\~no}(2025)}]{Redondo-Yuste:2025hlv}%
  \BibitemOpen
  \bibfield  {author} {\bibinfo {author} {\bibfnamefont {J.}~\bibnamefont
  {Redondo-Yuste}}\ and\ \bibinfo {author} {\bibfnamefont {A.}~\bibnamefont
  {C\'ardenas-Avenda\~no}},\ }\href@noop {} {\  (\bibinfo {year} {2025})},\
  \Eprint {http://arxiv.org/abs/2502.18643} {arXiv:2502.18643 [gr-qc]}
  \BibitemShut {NoStop}%
\bibitem [{\citenamefont {Collodel}\ and\ \citenamefont
  {Doneva}(2022)}]{Collodel:2022jly}%
  \BibitemOpen
  \bibfield  {author} {\bibinfo {author} {\bibfnamefont {L.~G.}\ \bibnamefont
  {Collodel}}\ and\ \bibinfo {author} {\bibfnamefont {D.~D.}\ \bibnamefont
  {Doneva}},\ }\href {\doibase 10.1103/PhysRevD.106.084057} {\bibfield
  {journal} {\bibinfo  {journal} {Phys. Rev. D}\ }\textbf {\bibinfo {volume}
  {106}},\ \bibinfo {pages} {084057} (\bibinfo {year} {2022})},\ \Eprint
  {http://arxiv.org/abs/2203.08203} {arXiv:2203.08203 [gr-qc]} \BibitemShut
  {NoStop}%
\bibitem [{\citenamefont {Evstafyeva}\ \emph {et~al.}(2023)\citenamefont
  {Evstafyeva}, \citenamefont {Rosca-Mead}, \citenamefont {Sperhake},\ and\
  \citenamefont {Bruegmann}}]{Evstafyeva:2023kfg}%
  \BibitemOpen
  \bibfield  {author} {\bibinfo {author} {\bibfnamefont {T.}~\bibnamefont
  {Evstafyeva}}, \bibinfo {author} {\bibfnamefont {R.}~\bibnamefont
  {Rosca-Mead}}, \bibinfo {author} {\bibfnamefont {U.}~\bibnamefont
  {Sperhake}}, \ and\ \bibinfo {author} {\bibfnamefont {B.}~\bibnamefont
  {Bruegmann}},\ }\href {\doibase 10.1103/PhysRevD.108.104064} {\bibfield
  {journal} {\bibinfo  {journal} {Phys. Rev. D}\ }\textbf {\bibinfo {volume}
  {108}},\ \bibinfo {pages} {104064} (\bibinfo {year} {2023})},\ \Eprint
  {http://arxiv.org/abs/2310.05200} {arXiv:2310.05200 [gr-qc]} \BibitemShut
  {NoStop}%
\bibitem [{\citenamefont {Baumgarte}\ and\ \citenamefont
  {Shapiro}(1998)}]{Baumgarte:1998te}%
  \BibitemOpen
  \bibfield  {author} {\bibinfo {author} {\bibfnamefont {T.~W.}\ \bibnamefont
  {Baumgarte}}\ and\ \bibinfo {author} {\bibfnamefont {S.~L.}\ \bibnamefont
  {Shapiro}},\ }\href {\doibase 10.1103/PhysRevD.59.024007} {\bibfield
  {journal} {\bibinfo  {journal} {Phys. Rev. D}\ }\textbf {\bibinfo {volume}
  {59}},\ \bibinfo {pages} {024007} (\bibinfo {year} {1998})},\ \bibinfo {note}
  {gr-qc/9810065}\BibitemShut {NoStop}%
\bibitem [{\citenamefont {Shibata}\ and\ \citenamefont
  {Nakamura}(1995)}]{Shibata:1995we}%
  \BibitemOpen
  \bibfield  {author} {\bibinfo {author} {\bibfnamefont {M.}~\bibnamefont
  {Shibata}}\ and\ \bibinfo {author} {\bibfnamefont {T.}~\bibnamefont
  {Nakamura}},\ }\href {\doibase 10.1103/PhysRevD.52.5428} {\bibfield
  {journal} {\bibinfo  {journal} {Phys. Rev. D}\ }\textbf {\bibinfo {volume}
  {52}},\ \bibinfo {pages} {5428} (\bibinfo {year} {1995})}\BibitemShut
  {NoStop}%
\bibitem [{\citenamefont {Alic}\ \emph {et~al.}(2012)\citenamefont {Alic},
  \citenamefont {Bona-Casas}, \citenamefont {Bona}, \citenamefont {Rezzolla},\
  and\ \citenamefont {Palenzuela}}]{Alic:2011gg}%
  \BibitemOpen
  \bibfield  {author} {\bibinfo {author} {\bibfnamefont {D.}~\bibnamefont
  {Alic}}, \bibinfo {author} {\bibfnamefont {C.}~\bibnamefont {Bona-Casas}},
  \bibinfo {author} {\bibfnamefont {C.}~\bibnamefont {Bona}}, \bibinfo {author}
  {\bibfnamefont {L.}~\bibnamefont {Rezzolla}}, \ and\ \bibinfo {author}
  {\bibfnamefont {C.}~\bibnamefont {Palenzuela}},\ }\href {\doibase
  10.1103/PhysRevD.85.064040} {\bibfield  {journal} {\bibinfo  {journal} {Phys.
  Rev. D}\ }\textbf {\bibinfo {volume} {85}},\ \bibinfo {pages} {064040}
  (\bibinfo {year} {2012})},\ \bibinfo {note} {arXiv:1106.2254
  [gr-qc]}\BibitemShut {NoStop}%
\bibitem [{\citenamefont {Cook}\ \emph {et~al.}(2016)\citenamefont {Cook},
  \citenamefont {Figueras}, \citenamefont {Kunesch}, \citenamefont {Sperhake},\
  and\ \citenamefont {Tunyasuvunakool}}]{Cook:2016soy}%
  \BibitemOpen
  \bibfield  {author} {\bibinfo {author} {\bibfnamefont {W.~G.}\ \bibnamefont
  {Cook}}, \bibinfo {author} {\bibfnamefont {P.}~\bibnamefont {Figueras}},
  \bibinfo {author} {\bibfnamefont {M.}~\bibnamefont {Kunesch}}, \bibinfo
  {author} {\bibfnamefont {U.}~\bibnamefont {Sperhake}}, \ and\ \bibinfo
  {author} {\bibfnamefont {S.}~\bibnamefont {Tunyasuvunakool}},\ }\href
  {\doibase 10.1142/S0218271816410133} {\bibfield  {journal} {\bibinfo
  {journal} {Int. J. Mod. Phys. D}\ }\textbf {\bibinfo {volume} {25}},\
  \bibinfo {pages} {1641013} (\bibinfo {year} {2016})},\ \bibinfo {note}
  {arXiv:1603.00362 [gr-qc]}\BibitemShut {NoStop}%
\bibitem [{new(2024)}]{newexozvezda}%
  \BibitemOpen
  \href@noop {} {\enquote {\bibinfo {title} {{ExoZvezda code;\\
  https://github.com/GRTLCollaboration/ExoZvezda}},}\ }\bibinfo {howpublished}
  {GitHub repository} (\bibinfo {year} {2024})\BibitemShut {NoStop}%
\bibitem [{\citenamefont {Radia}\ \emph {et~al.}(2022)\citenamefont {Radia},
  \citenamefont {Sperhake}, \citenamefont {Drew}, \citenamefont {Clough},
  \citenamefont {Figueras}, \citenamefont {Lim}, \citenamefont {Ripley},
  \citenamefont {Aurrekoetxea}, \citenamefont {Fran\c{c}a},\ and\ \citenamefont
  {Helfer}}]{Radia:2021smk}%
  \BibitemOpen
  \bibfield  {author} {\bibinfo {author} {\bibfnamefont {M.}~\bibnamefont
  {Radia}}, \bibinfo {author} {\bibfnamefont {U.}~\bibnamefont {Sperhake}},
  \bibinfo {author} {\bibfnamefont {A.}~\bibnamefont {Drew}}, \bibinfo {author}
  {\bibfnamefont {K.}~\bibnamefont {Clough}}, \bibinfo {author} {\bibfnamefont
  {P.}~\bibnamefont {Figueras}}, \bibinfo {author} {\bibfnamefont {E.~A.}\
  \bibnamefont {Lim}}, \bibinfo {author} {\bibfnamefont {J.~L.}\ \bibnamefont
  {Ripley}}, \bibinfo {author} {\bibfnamefont {J.~C.}\ \bibnamefont
  {Aurrekoetxea}}, \bibinfo {author} {\bibfnamefont {T.}~\bibnamefont
  {Fran\c{c}a}}, \ and\ \bibinfo {author} {\bibfnamefont {T.}~\bibnamefont
  {Helfer}},\ }\href {\doibase 10.1088/1361-6382/ac6fa9} {\bibfield  {journal}
  {\bibinfo  {journal} {Class. Quant. Grav.}\ }\textbf {\bibinfo {volume}
  {39}},\ \bibinfo {pages} {135006} (\bibinfo {year} {2022})},\ \Eprint
  {http://arxiv.org/abs/2112.10567} {arXiv:2112.10567 [gr-qc]} \BibitemShut
  {NoStop}%
\bibitem [{\citenamefont {Andrade}\ \emph {et~al.}(2021)\citenamefont {Andrade}
  \emph {et~al.}}]{Andrade:2021rbd}%
  \BibitemOpen
  \bibfield  {author} {\bibinfo {author} {\bibfnamefont {T.}~\bibnamefont
  {Andrade}} \emph {et~al.},\ }\href {\doibase 10.21105/joss.03703} {\bibfield
  {journal} {\bibinfo  {journal} {J. Open Source Softw.}\ }\textbf {\bibinfo
  {volume} {6}},\ \bibinfo {pages} {3703} (\bibinfo {year} {2021})},\ \Eprint
  {http://arxiv.org/abs/2201.03458} {arXiv:2201.03458 [gr-qc]} \BibitemShut
  {NoStop}%
\bibitem [{\citenamefont {Adams}\ \emph {et~al.}(2019)\citenamefont {Adams}
  \emph {et~al.}}]{chombo}%
  \BibitemOpen
  \bibfield  {author} {\bibinfo {author} {\bibfnamefont {M.}~\bibnamefont
  {Adams}} \emph {et~al.},\ }\href@noop {} {\emph {\bibinfo {title} {{Chombo
  Software Package for AMR Applications - Design Document}}}},\ \bibinfo {type}
  {Tech. Rep.}\ \bibinfo {number} {{LBNL}-6616{E}}\ (\bibinfo  {institution}
  {Lawrence Berkeley National Laboratory},\ \bibinfo {year} {2019})\BibitemShut
  {NoStop}%
\bibitem [{\citenamefont {Sperhake}(2007)}]{Sperhake:2006cy}%
  \BibitemOpen
  \bibfield  {author} {\bibinfo {author} {\bibfnamefont {U.}~\bibnamefont
  {Sperhake}},\ }\href {\doibase 10.1103/PhysRevD.76.104015} {\bibfield
  {journal} {\bibinfo  {journal} {Phys. Rev. D}\ }\textbf {\bibinfo {volume}
  {76}},\ \bibinfo {pages} {104015} (\bibinfo {year} {2007})},\ \bibinfo {note}
  {gr-qc/0606079}\BibitemShut {NoStop}%
\bibitem [{\citenamefont {{Allen, G. and Goodale, T. and Mass\'o, J. and
  Seidel, E.}}(1999)}]{Allen:1999}%
  \BibitemOpen
  \bibfield  {author} {\bibinfo {author} {\bibnamefont {{Allen, G. and Goodale,
  T. and Mass\'o, J. and Seidel, E.}}},\ }in\ \href@noop {} {\emph {\bibinfo
  {booktitle} {{Proceedings of Eighth IEEE International Symposium on High
  Performance Distributed Computing, HPDC-8, Redondo Beach, 1999}}}}\ (\bibinfo
   {publisher} {{IEEE Press}},\ \bibinfo {address} {{Piscataway, New Jersey,
  United States}},\ \bibinfo {year} {1999})\BibitemShut {NoStop}%
\bibitem [{\citenamefont {Schnetter}\ \emph {et~al.}(2004)\citenamefont
  {Schnetter}, \citenamefont {Hawley},\ and\ \citenamefont
  {Hawke}}]{Schnetter:2003rb}%
  \BibitemOpen
  \bibfield  {author} {\bibinfo {author} {\bibfnamefont {E.}~\bibnamefont
  {Schnetter}}, \bibinfo {author} {\bibfnamefont {S.~H.}\ \bibnamefont
  {Hawley}}, \ and\ \bibinfo {author} {\bibfnamefont {I.}~\bibnamefont
  {Hawke}},\ }\href {\doibase 10.1088/0264-9381/21/6/014} {\bibfield  {journal}
  {\bibinfo  {journal} {Class. Quant. Grav.}\ }\textbf {\bibinfo {volume}
  {21}},\ \bibinfo {pages} {1465} (\bibinfo {year} {2004})},\ \bibinfo {note}
  {gr-qc/0310042}\BibitemShut {NoStop}%
\bibitem [{\citenamefont {Alic}\ \emph {et~al.}(2013)\citenamefont {Alic},
  \citenamefont {Kastaun},\ and\ \citenamefont {Rezzolla}}]{Alic:2013xsa}%
  \BibitemOpen
  \bibfield  {author} {\bibinfo {author} {\bibfnamefont {D.}~\bibnamefont
  {Alic}}, \bibinfo {author} {\bibfnamefont {W.}~\bibnamefont {Kastaun}}, \
  and\ \bibinfo {author} {\bibfnamefont {L.}~\bibnamefont {Rezzolla}},\ }\href
  {\doibase 10.1103/PhysRevD.88.064049} {\bibfield  {journal} {\bibinfo
  {journal} {Phys. Rev. D}\ }\textbf {\bibinfo {volume} {88}},\ \bibinfo
  {pages} {064049} (\bibinfo {year} {2013})},\ \Eprint
  {http://arxiv.org/abs/arXiv:1307.7391 [gr-qc]} {arXiv:1307.7391 [gr-qc]}
  \BibitemShut {NoStop}%
\bibitem [{\citenamefont {Gleiser}\ and\ \citenamefont
  {Watkins}(1989)}]{Gleiser:1988ih}%
  \BibitemOpen
  \bibfield  {author} {\bibinfo {author} {\bibfnamefont {M.}~\bibnamefont
  {Gleiser}}\ and\ \bibinfo {author} {\bibfnamefont {R.}~\bibnamefont
  {Watkins}},\ }\href {\doibase 10.1016/0550-3213(89)90627-5} {\bibfield
  {journal} {\bibinfo  {journal} {Nucl. Phys. B}\ }\textbf {\bibinfo {volume}
  {319}},\ \bibinfo {pages} {733} (\bibinfo {year} {1989})}\BibitemShut
  {NoStop}%
\bibitem [{\citenamefont {Kain}(2021)}]{Kain:2021rmk}%
  \BibitemOpen
  \bibfield  {author} {\bibinfo {author} {\bibfnamefont {B.}~\bibnamefont
  {Kain}},\ }\href {\doibase 10.1103/PhysRevD.103.123003} {\bibfield  {journal}
  {\bibinfo  {journal} {Phys. Rev. D}\ }\textbf {\bibinfo {volume} {103}},\
  \bibinfo {pages} {123003} (\bibinfo {year} {2021})},\ \Eprint
  {http://arxiv.org/abs/2106.01740} {arXiv:2106.01740 [gr-qc]} \BibitemShut
  {NoStop}%
\bibitem [{\citenamefont {Rosa}\ \emph {et~al.}(2023)\citenamefont {Rosa},
  \citenamefont {Macedo},\ and\ \citenamefont {Rubiera-Garcia}}]{Rosa:2023qcv}%
  \BibitemOpen
  \bibfield  {author} {\bibinfo {author} {\bibfnamefont {J.~L.}\ \bibnamefont
  {Rosa}}, \bibinfo {author} {\bibfnamefont {C.~F.~B.}\ \bibnamefont {Macedo}},
  \ and\ \bibinfo {author} {\bibfnamefont {D.}~\bibnamefont {Rubiera-Garcia}},\
  }\href {\doibase 10.1103/PhysRevD.108.044021} {\bibfield  {journal} {\bibinfo
   {journal} {Phys. Rev. D}\ }\textbf {\bibinfo {volume} {108}},\ \bibinfo
  {pages} {044021} (\bibinfo {year} {2023})},\ \Eprint
  {http://arxiv.org/abs/2303.17296} {arXiv:2303.17296 [gr-qc]} \BibitemShut
  {NoStop}%
\bibitem [{\citenamefont {Mayerson}()}]{newMayerson:2024fgh}%
  \BibitemOpen
  \bibfield  {author} {\bibinfo {author} {\bibfnamefont {D.~R.}\ \bibnamefont
  {Mayerson}},\ }\href@noop {} {\enquote {\bibinfo {title} {{FOORT: Flexible
  Object-Oriented Ray Tracing; https://github.com/drmayerson/FOORT}},}\
  }\BibitemShut {NoStop}%
\bibitem [{\citenamefont {Mayerson}\ \emph {et~al.}()\citenamefont {Mayerson},
  \citenamefont {Bacchini},\ and\ \citenamefont {Staelens}}]{Mayerson:2025foo}%
  \BibitemOpen
  \bibfield  {author} {\bibinfo {author} {\bibfnamefont {D.~R.}\ \bibnamefont
  {Mayerson}}, \bibinfo {author} {\bibfnamefont {F.}~\bibnamefont {Bacchini}},
  \ and\ \bibinfo {author} {\bibfnamefont {S.~J.}\ \bibnamefont {Staelens}},\
  }\href@noop {} {\ }\bibinfo {note} {(in prep.)}\BibitemShut {NoStop}%
\end{thebibliography}

\begin{thebibliography}{9}%
\makeatletter
\providecommand \@ifxundefined [1]{%
 \@ifx{#1\undefined}
}%
\providecommand \@ifnum [1]{%
 \ifnum #1\expandafter \@firstoftwo
 \else \expandafter \@secondoftwo
 \fi
}%
\providecommand \@ifx [1]{%
 \ifx #1\expandafter \@firstoftwo
 \else \expandafter \@secondoftwo
 \fi
}%
\providecommand \natexlab [1]{#1}%
\providecommand \enquote  [1]{``#1''}%
\providecommand \bibnamefont  [1]{#1}%
\providecommand \bibfnamefont [1]{#1}%
\providecommand \citenamefont [1]{#1}%
\providecommand \href@noop [0]{\@secondoftwo}%
\providecommand \href [0]{\begingroup \@sanitize@url \@href}%
\providecommand \@href[1]{\@@startlink{#1}\@@href}%
\providecommand \@@href[1]{\endgroup#1\@@endlink}%
\providecommand \@sanitize@url [0]{\catcode `\\12\catcode `\$12\catcode
  `\&12\catcode `\#12\catcode `\^12\catcode `\_12\catcode `\%12\relax}%
\providecommand \@@startlink[1]{}%
\providecommand \@@endlink[0]{}%
\providecommand \url  [0]{\begingroup\@sanitize@url \@url }%
\providecommand \@url [1]{\endgroup\@href {#1}{\urlprefix }}%
\providecommand \urlprefix  [0]{URL }%
\providecommand \Eprint [0]{\href }%
\providecommand \doibase [0]{https://doi.org/}%
\providecommand \selectlanguage [0]{\@gobble}%
\providecommand \bibinfo  [0]{\@secondoftwo}%
\providecommand \bibfield  [0]{\@secondoftwo}%
\providecommand \translation [1]{[#1]}%
\providecommand \BibitemOpen [0]{}%
\providecommand \bibitemStop [0]{}%
\providecommand \bibitemNoStop [0]{.\EOS\space}%
\providecommand \EOS [0]{\spacefactor3000\relax}%
\providecommand \BibitemShut  [1]{\csname bibitem#1\endcsname}%
\let\auto@bib@innerbib\@empty
\bibitem [{\citenamefont {Kain}(2021)}]{SMKain:2021rmk}%
  \BibitemOpen
  \bibfield  {author} {\bibinfo {author} {\bibfnamefont {B.}~\bibnamefont
  {Kain}},\ }\bibfield  {title} {\bibinfo {title} {{Boson stars and their
  radial oscillations}},\ }\href {https://doi.org/10.1103/PhysRevD.103.123003}
  {\bibfield  {journal} {\bibinfo  {journal} {Phys. Rev. D}\ }\textbf {\bibinfo
  {volume} {103}},\ \bibinfo {pages} {123003} (\bibinfo {year} {2021})},\
  \Eprint {https://arxiv.org/abs/2106.01740} {arXiv:2106.01740 [gr-qc]}
  \BibitemShut {NoStop}%
\bibitem [{\citenamefont {Gleiser}\ and\ \citenamefont
  {Watkins}(1989)}]{SMGleiser:1988ih}%
  \BibitemOpen
  \bibfield  {author} {\bibinfo {author} {\bibfnamefont {M.}~\bibnamefont
  {Gleiser}}\ and\ \bibinfo {author} {\bibfnamefont {R.}~\bibnamefont
  {Watkins}},\ }\bibfield  {title} {\bibinfo {title} {{Gravitational Stability
  of Scalar Matter}},\ }\href {https://doi.org/10.1016/0550-3213(89)90627-5}
  {\bibfield  {journal} {\bibinfo  {journal} {Nucl. Phys. B}\ }\textbf
  {\bibinfo {volume} {319}},\ \bibinfo {pages} {733} (\bibinfo {year}
  {1989})}\BibitemShut {NoStop}%
\bibitem [{\citenamefont {Ge}\ \emph {et~al.}(2024)\citenamefont {Ge},
  \citenamefont {Lim}, \citenamefont {Sperhake}, \citenamefont {Evstafyeva},
  \citenamefont {Cors}, \citenamefont {de~Jong}, \citenamefont {Croft},\ and\
  \citenamefont {Helfer}}]{SMGe:2024itl}%
  \BibitemOpen
  \bibfield  {author} {\bibinfo {author} {\bibfnamefont {B.-X.}\ \bibnamefont
  {Ge}}, \bibinfo {author} {\bibfnamefont {E.~A.}\ \bibnamefont {Lim}},
  \bibinfo {author} {\bibfnamefont {U.}~\bibnamefont {Sperhake}}, \bibinfo
  {author} {\bibfnamefont {T.}~\bibnamefont {Evstafyeva}}, \bibinfo {author}
  {\bibfnamefont {D.}~\bibnamefont {Cors}}, \bibinfo {author} {\bibfnamefont
  {E.}~\bibnamefont {de~Jong}}, \bibinfo {author} {\bibfnamefont
  {R.}~\bibnamefont {Croft}},\ and\ \bibinfo {author} {\bibfnamefont
  {T.}~\bibnamefont {Helfer}},\ }\bibfield  {title} {\bibinfo {title} {{Hair is
  complicated: Gravitational waves from stable and unstable boson-star
  mergers}},\ }\href@noop {} {\  (\bibinfo {year} {2024})},\ \Eprint
  {https://arxiv.org/abs/2410.23839} {arXiv:2410.23839 [gr-qc]} \BibitemShut
  {NoStop}%
\bibitem [{\citenamefont {Cunha}\ \emph {et~al.}(2023)\citenamefont {Cunha},
  \citenamefont {Herdeiro}, \citenamefont {Radu},\ and\ \citenamefont
  {Sanchis-Gual}}]{SMCunha:2022gde}%
  \BibitemOpen
  \bibfield  {author} {\bibinfo {author} {\bibfnamefont {P.~V.~P.}\
  \bibnamefont {Cunha}}, \bibinfo {author} {\bibfnamefont {C.}~\bibnamefont
  {Herdeiro}}, \bibinfo {author} {\bibfnamefont {E.}~\bibnamefont {Radu}},\
  and\ \bibinfo {author} {\bibfnamefont {N.}~\bibnamefont {Sanchis-Gual}},\
  }\bibfield  {title} {\bibinfo {title} {{Exotic Compact Objects and the Fate
  of the Light-Ring Instability}},\ }\href
  {https://doi.org/10.1103/PhysRevLett.130.061401} {\bibfield  {journal}
  {\bibinfo  {journal} {Phys. Rev. Lett.}\ }\textbf {\bibinfo {volume} {130}},\
  \bibinfo {pages} {061401} (\bibinfo {year} {2023})},\ \Eprint
  {https://arxiv.org/abs/2207.13713} {arXiv:2207.13713 [gr-qc]} \BibitemShut
  {NoStop}%
\bibitem [{\citenamefont {Mayerson}()}]{SMnewMayerson:2024fgh}%
  \BibitemOpen
  \bibfield  {author} {\bibinfo {author} {\bibfnamefont {D.~R.}\ \bibnamefont
  {Mayerson}},\ }\href@noop {} {\enquote {\bibinfo {title} {{FOORT: Flexible
  Object-Oriented Ray Tracing;\\ https://github.com/drmayerson/FOORT}},}\
  }\BibitemShut {NoStop}%
\bibitem [{\citenamefont {Mayerson}\ \emph {et~al.}()\citenamefont {Mayerson},
  \citenamefont {Bacchini},\ and\ \citenamefont {Staelens}}]{SMMayerson:2025foo}%
  \BibitemOpen
  \bibfield  {author} {\bibinfo {author} {\bibfnamefont {D.~R.}\ \bibnamefont
  {Mayerson}}, \bibinfo {author} {\bibfnamefont {F.}~\bibnamefont {Bacchini}},\
  and\ \bibinfo {author} {\bibfnamefont {S.~J.}\ \bibnamefont {Staelens}},\
  }\bibfield  {title} {\bibinfo {title} {{FOORT: Flexible Object-Oriented Ray
  Tracing}},\ }\href@noop {} {\ }\bibinfo {note} {(in prep.)}\BibitemShut
  {NoStop}%
\bibitem [{\citenamefont {Gralla}\ \emph {et~al.}(2020)\citenamefont {Gralla},
  \citenamefont {Lupsasca},\ and\ \citenamefont {Marrone}}]{SMGralla:2020sha}%
  \BibitemOpen
  \bibfield  {author} {\bibinfo {author} {\bibfnamefont {S.~E.}\ \bibnamefont
  {Gralla}}, \bibinfo {author} {\bibfnamefont {A.}~\bibnamefont {Lupsasca}},\
  and\ \bibinfo {author} {\bibfnamefont {D.~P.}\ \bibnamefont {Marrone}},\
  }\bibfield  {title} {\bibinfo {title} {The shape of the black hole photon
  ring: A precise test of strong-field general relativity},\ }\href
  {https://doi.org/10.1103/PhysRevD.102.124004} {\bibfield  {journal} {\bibinfo
   {journal} {Physical Review D}\ }\textbf {\bibinfo {volume} {102}},\ \bibinfo
  {pages} {124004} (\bibinfo {year} {2020})}\BibitemShut {NoStop}%
\bibitem [{\citenamefont {Pu}\ \emph {et~al.}(2016)\citenamefont {Pu},
  \citenamefont {Akiyama},\ and\ \citenamefont {Asada}}]{SMPu:2016flo}%
  \BibitemOpen
  \bibfield  {author} {\bibinfo {author} {\bibfnamefont {H.-Y.}\ \bibnamefont
  {Pu}}, \bibinfo {author} {\bibfnamefont {K.}~\bibnamefont {Akiyama}},\ and\
  \bibinfo {author} {\bibfnamefont {K.}~\bibnamefont {Asada}},\ }\bibfield
  {title} {\bibinfo {title} {The effects of accretion flow dynamics on the
  black hole shadow of {S}agittarius {A}*},\ }\href
  {https://doi.org/10.3847/0004-637X/831/1/4} {\bibfield  {journal} {\bibinfo
  {journal} {The Astrophysical Journal}\ }\textbf {\bibinfo {volume} {831}},\
  \bibinfo {pages} {4} (\bibinfo {year} {2016})}\BibitemShut {NoStop}%
\bibitem [{\citenamefont {C\'ardenas-Avenda\~no}\ \emph
  {et~al.}(2023)\citenamefont {C\'ardenas-Avenda\~no}, \citenamefont
  {Lupsasca},\ and\ \citenamefont {Zhu}}]{SMCardenasAvendano:2023aar}%
  \BibitemOpen
  \bibfield  {author} {\bibinfo {author} {\bibfnamefont {A.}~\bibnamefont
  {C\'ardenas-Avenda\~no}}, \bibinfo {author} {\bibfnamefont {A.}~\bibnamefont
  {Lupsasca}},\ and\ \bibinfo {author} {\bibfnamefont {H.}~\bibnamefont
  {Zhu}},\ }\bibfield  {title} {\bibinfo {title} {Adaptive analytical ray
  tracing of black hole photon rings},\ }\href
  {https://doi.org/10.1103/PhysRevD.107.043030} {\bibfield  {journal} {\bibinfo
   {journal} {Phys. Rev. D}\ }\textbf {\bibinfo {volume} {107}},\ \bibinfo
  {pages} {043030} (\bibinfo {year} {2023})}\BibitemShut {NoStop}%
\end{thebibliography}
\end{document}